# Strategies to Enhance ZnO Nanogenerator Performance via Thermal-Annealing and Cryo-Cooling

Guylaine Poulin-Vittrant, Kevin Nadaud, Chandrakanth Reddy Chandraiahgari and Daniel Alquier

GREMAN UMR 7347, CNRS, Université de Tours, INSA-CVL, 16 rue Pierre et Marie Curie, BP 7155, 37071 TOURS cedex 2, FRANCE
Correspondence: guylaine.poulin-vittrant@univ-tours.fr

**Abstract:** Piezoelectric capacitive NanoGenerators (NG) based on vertically grown crystalline zinc oxide nanowires (ZnO-NWs) have been fabricated using a low-cost and scalable hydrothermal method on gold-coated silicon substrates, which served as both a seed layer and a conductive bottom electrode. Morphological and structural characterizations demonstrate that the obtained ZnO NWs are dense, uniformly distributed, vertically well aligned and exhibit good crystal quality. The piezoelectric NG consists of ZnO NWs grown on a gold-coated silicon substrate, parylene-C matrix, titanium/aluminium top electrode and poly(dimethylsiloxane) (PDMS) encapsulating layer. In order to enhance the NG performances, which is the main goal of this study, two distinctly different post-growth treatments, namely thermal annealing in ambient air and cryo-cooling by immersion in liquid nitrogen, are applied and their effect studied. Achieving the high performance of NG via the combination of high-quality NWs growth and subsequent post-growth treatment is presented. Superior global performance of NG has been observed with a post-treatment of cryo-cooling for an optimum duration compared to the thermal annealing signifies the simplicity and novelty of the work. The proposed strategies highlight the role of post-growth treatments towards the fabrication of high-performance functional NG to be incorporated into future smart objects.

**Keywords:** Piezoelectric nanogenerators; zinc oxide; hydrothermal method; post-growth treatment; thermal annealing; liquid nitrogen.

## 1. Introduction

Nanotechnology enabled energy harvesting devices have emerged fifteen years ago and have arisen intense research since then [1,2]. Among the various available renewable energy sources, mechanical energy is ubiquitous and can be a potential source of green energy to power the electronic devices [3]. Remarkable achievements have been made in the research and development of piezoelectric and sensing devices based on functional piezoelectric materials, ever since the first piezoelectric NanoGenerators (NG) was developed in 2006 by Wang et al. that uses zinc oxide nanowires (ZnO-NWs) to convert mechanical energy into electrical energy [4]. Piezoelectric NGs are rapidly emerging with promising abilities for harvesting random mechanical energy into electric energy through nanometer-scale piezoelectric materials. ZnO is an important functional material due to its desirable piezoelectric semiconducting, geometric versatility, and excellent biocompatible properties. It can be grown easily in NWs at low temperature (85 °C) on several substrates using a well-established facile HydroThermal (HT) method [5]. These distinctive features of ZnO make it a qualified candidate for fabricating NG. In recent years, vertically aligned ZnO-NWs array-based NG is one of the dominant designs developed to harvest mechanical energy using piezoelectric nanostructures [6,7]. These developments have as a goal to harvest ambient mechanical energy and then to utilize the converted energy to operate electronic devices, by ensuring their energy autonomy. Moreover, due to their small size, NG can be effectively integrated with other nano/micro-scale functional devices to build self-powered systems [8]. In this context, NG which exploits ambient energy sources to power the micro/nano-systems has been proposed for the development of self-powered electronics.

The nanostructures of ZnO can generate a piezoelectric potential of a few volts when subjected to a mechanical deformation [6,8]. Nevertheless, free charge carriers exist in ZnO and are suspected to screen some part of the generated piezoelectric potential, resulting in a lower output power generation by the NG [9]. It was shown that, in nanostructured ZnO, native point defects play a central role in defining the electronic device performances [10]. Such disturbing phenomena occurring in semiconducting NWs limit the performance of NG and eventually, there is still room for improvement before reaching industrial market [3]. Therefore, it signifies the need for novel strategies to enhance the performance of NGs. To solve this problem, several approaches including NG design, improvements in the integration, chemical doping, micro/nanostructure morphology of ZnO-NWs have already been investigated [11-18]. Most of these optimization techniques reported with the aim of improving the NG performance focus on either re-designing the materials or improving the device structures.

To date, very few reports are available on enhancing the performance of NGs via post-growth treatments. Post-growth treatment is just an extra step that should take place subsequently after the growth of NWs without altering any other existing device fabrication process. Thermal annealing is one such simple low-cost approach widely adopted albeit not suitable for flexible polymer substrates. For instance, it has been shown that the performance of solar cells, UV emission in ZnO films and piezoelectric constant in AlN films was enhanced upon thermal annealing and ascribed to the improved morphology, crystallinity, and relaxation of the internal compressive stress respectively [19-21]. In a study by our group, the high carrier concentration in ZnO-NWs was greatly suppressed by thermal annealing in ambient air at ~450 °C [22]. Recently, Fortunato et al. demonstrated that quenching in liquid nitrogen greatly improved the crystallinity and piezoelectric constant in PVDF films [23]. Liquid nitrogen (LN) is a colorless cryogenic fluid at an extremely low temperature of -196 °C and often used in space and industrial applications. Therefore, herein we have investigated the two different post-growth treatments, namely thermal annealing in ambient air and cryo-cooling by immersion in LN, owing to their vital role in enhancing the functional properties. Cryo-cooling is a complementary approach to the high-temperature thermal annealing and is a novel post-growth treatment being reported for the first time for ZnO-NWs.

Therefore, this work is aimed at investigating the two distinctly different post-growth treatments (thermal-annealing and cryo-cooling) by assessing the ZnO-NWs morphological and crystalline properties, and eventually the performance of functional NGs is reported.

**2. Materials and Methods**

*2.1. Materials*

All the chemicals used for the ZnO-NW growth were of reagent-grade and were used as received: zinc nitrate hexahydrate ($Zn(NO_3)_2 \cdot 6H_2O$, Sigma-Aldrich, ACS reagent, ≥99%), hexamethylenetetramine ($C_6H_{12}N_4$, Sigma-Aldrich, ACS reagent, ≥99%, solution from Saint-Quentin Fallavier, France), ammonium hydroxide (ammonia) ($NH_4OH$, 29%, solution from KMG Ultra-Pure Chemicals, Saint Fromond, France), silver paint (TED PELLA, USA), and deionized (DI) water (16 MΩ cm). The substrate cleaning was carried out with hydrofluoric acid HF (50%), hydrogen peroxide $H_2O_2$ (30%), and sulfuric acid $H_2SO_4$ (96%), which were supplied by KMG Ultra-Pure Chemicals (Saint Fromond, France). The deposition of a metallic layer was done with physical vapor deposition (PVD) equipment (Plassys MP 650 S, Marolles-en-Hurepoix, France). A tubular furnace (Thermolyne79300, Dubuque, IA, USA) was used for the annealing treatment. A stainless-steel autoclave (from Parr Instrument Company, Moline, IL, USA) was used to perform the growth reaction.

*2.2. Growth of ZnO Nanowires*

The ZnO-NWs were grown by a facile and low-temperature HT growth process on (100) oriented Si wafers according to the detailed procedure and growth mechanism described in our previous work [24, 25]. Briefly, a 500 μm thick n-type Si substrate with 2×2 $cm^2$ area was first cleaned in piranha solution (1:1, $H_2SO_4$ and $H_2O_2$) for 10 min followed by a 2-min dip in hydrofluoric acid (1:1, HF and $H_2O$) to remove the thin oxide and, finally, rinsing in DI water. This cleaning step was followed by drying under nitrogen gas, and a final baking step was performed at 200 °C for 15 min to remove any adsorbed moisture prior to the metal deposition. A gold layer (~ 200 nm) was then deposited by direct current sputtering technique (500 W, 5 mTorr pressure in argon atmosphere) at room temperature, to serve as bottom electrode towards the subsequent assembly of the NG. To improve the adhesion between gold and silicon, a layer of titanium (~100 nm) was deposited using the same technique.

An equimolar growth solution was prepared by dissolving 100 mM of zinc nitrate hexahydrate and hexamine in 90 ml of deionized (DI) water. Ammonia (30 mM) was introduced in the growth solution. Our group has employed ammonia for the single-step growth of ZnO-NWs on gold surfaces as it simultaneously controls the NW density (defined as the number of NWs per unit area) and free charge density [24-25]. The solution was magnetically stirred at room temperature yielding a clear solution. The nutrient solution was then transferred into a Teflon flask which was sealed in a stainless-steel autoclave and placed inside a preheated convection oven maintained at 85 °C for 6 h for the growth reaction. During the growth, three gold deposited Si substrates were immersed facing down, with a ~60° slope against the walls. After the growth, the autoclave was removed from the oven and cooled down naturally. The substrates were then thoroughly rinsed with flowing DI water and dried under $N_2$ gas flow.

*2.3. Post-Growth Treatment*

Two distinctly different post-growth conditions namely thermal-annealing and cryo-cooling were applied. For high-temperature thermal annealing in air, the samples were placed in a preheated horizontal quartz tubular furnace and annealing was performed at two different temperatures, 350 °C for 90 min and 450 °C for 30 min, selected according to our previous work as they suppress the high concentration of excess free carriers [22]. For low-temperature cooling, as shown on Figure 1, the samples were immersed in a Teflon beaker filled with LN for two different durations, 15 min and 30 min, the samples hereinafter are denoted as LN15 and LN30 respectively. After immersion in LN, the samples are dried under $N_2$ gas flow.

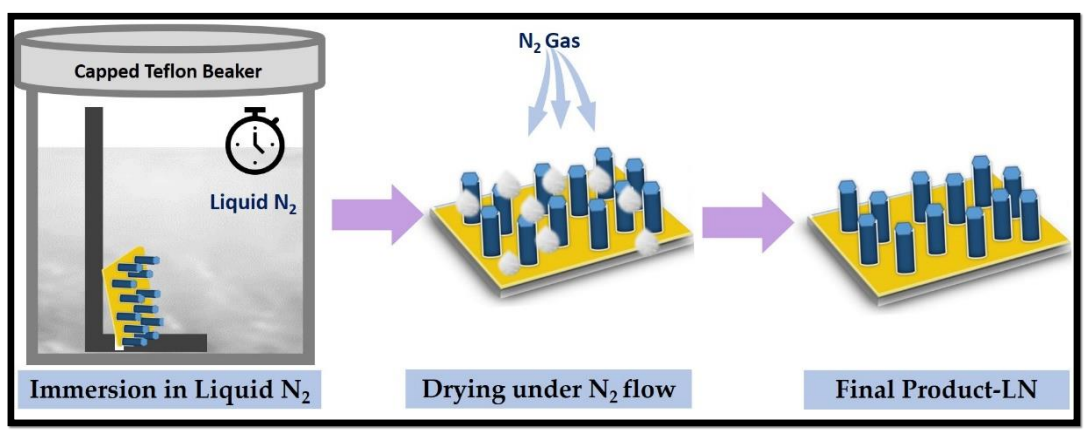

**Figure 1.** Schematic representation of cryo-cooling treatment for the ZnO-NWs

*2.4. NG Device Fabrication*

Fully functional NGs were assembled after the ZnO-NW growth subjected or not to the respective post-growth conditions. The scheme of the NG device fabrication process is depicted in Figure 2. First, a thin Parylene-C layer (~500 nm) was deposited over NWs and then 400 nm/100 nm thick Al/Ti layers were evaporated on the top surface to define the active working area 1.5 cm$^2$ of the devices. The role of Parylene-C was to guarantee the separation between bottom Au and top Al contacts and thereby, to serve as dielectric material to make the device a capacitive structure. Afterward, metal wires were bonded using the conductive silver paste. Finally, the NGs were encapsulated in thick PDMS to minimize degradation during testing.

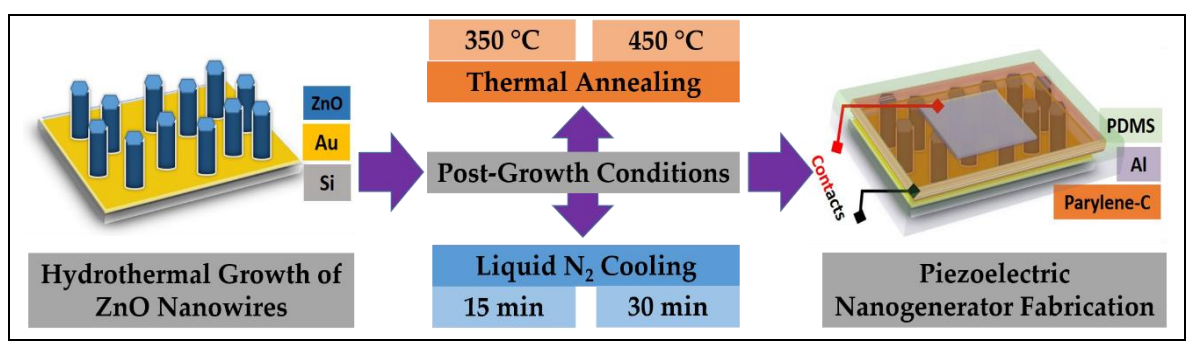

**Figure 2.** Schematic illustration of the NG device fabrication subjected or not to various post-growth treatments.

*2.5. Characterizations*

The surface morphology of ZnO-NWs was observed using a SEM (Hitachi S-4160). Samples were imaged without any metal coating, with an accelerating voltage of 10 keV (and the working distance was 13 mm). The structural characterization and phase identification were performed by a parallel beam diffractometer (BrukerAXS D8 discover, Karlsruhe, Germany) with a step size of 0.02° and a scan speed of 4 sec. XRD data were collected at room temperature, using Cu Kα radiation in a 2θ angular range from 20° to 60°.

Electrical NGs were characterized by a custom-built test-bench [26]. The performance measurement set-up was designed in order to apply a low-frequency compression force up to 13 N in the range of 1-10 Hz, like the targeted mechanical sources. At the output of the NG, a variable resistance from 1kΩ to 128 MΩ was connected in order to sweep the resistive load and find the load which maximize the harvested power, and the voltage was measured via a high input impedance double buffer circuit [27]. The output response of an energy harvester generally includes the average output voltage/electric potential ($V_{RMS}$), the average current ($I_{RMS}$) and the average power ($P_{AV}$) vs. load resistance ($R_L$). The measurements were performed under different loads, and for two amplitudes, 3 N and 6 N, of compression force at a frequency of 5 Hz. The average value of output responses of three devices, prepared for each type, is reported.

**3. Results and Discussions**

*3.1. Morphology analysis by FE-SEM*

The surface morphology observed by FE-SEM in Figure 3 shows that the produced nanostructures have rod-like uniform morphology with hexagonal structure at the tip surface and grown along their crystalline c-axis. The average diameter and length of NWs are 210 nm and 0.92 μm respectively. Morphology of the samples was not affected upon either type of post-growth treatments (images not shown, hereby).

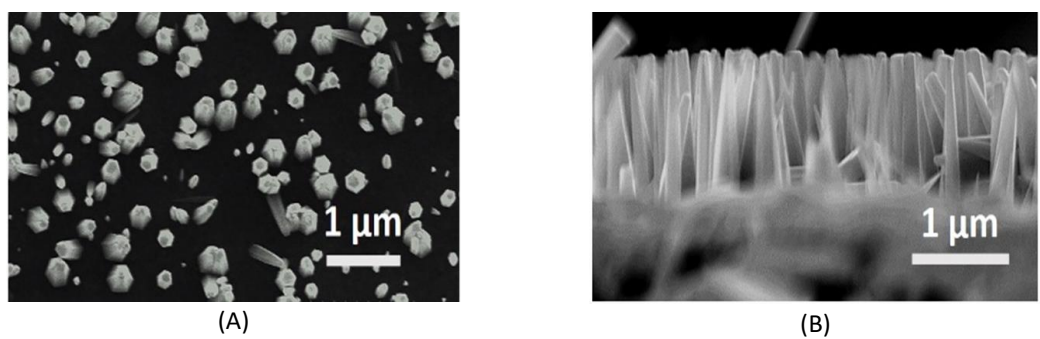

(A)    (B)

**Figure 3.** FE-SEM images of ZnO-NWs grown by HT method : (A) top view; (B) cross section.

*3.2. Crystalline structure analysis by XRD*

The XRD spectra recorded in the 2θ range 20°- 60° has only two peaks at 34.42° and 38.18°, which indicate the highly crystalline nature of the materials. According to the Joint Committee on Powder Diffraction Standards (JCPDS file Nr 005-0664), the 2θ peak at 34.42° is associated to (0 0 2) planes of wurtzite hexagonal structural phase of ZnO, predominantly grew along the c-axis. Furthermore, the 2θ peak at 38.18° is associated to (1 1 1) planes of Au (JCPDS File Nr 004-0784), and is due to the substrate deposited with gold prior to the growth. Therefore, herein the presented spectra are focused on the 34.2°-34.6° range to follow the evolution of 2θ peak at (0 0 2) measured on the different ZnO-NWs samples. Figure 4 shows the diffraction patterns of the ZnO-NWs subjected to various post-growth conditions.

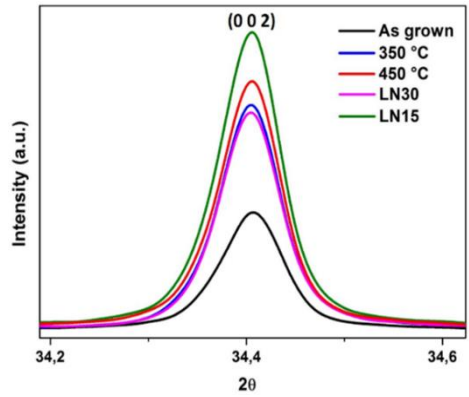

**Figure 4.** XRD patterns of ZnO-NWs subjected to various post-growth treatments.

The crystallite sizes (D) of the ZnO-NWs were calculated using the following Scherrer equation, where $\lambda$, $\theta$, and $\beta$ are the X-ray wavelength (0.154 nm), Bragg's diffraction angle, and full-width at half-maximum (FWHM) of the ZnO (002) diffraction peak, respectively.

$$D = \frac{0.9\lambda}{\beta \cos\theta} \quad (1)$$

The lattice constant *c* was calculated by the following equation, where h, k, and l are crystal Miller indices, and *d* is the interplanar spacing, respectively.

$$\frac{1}{d^2_{(hkl)}} = \frac{4}{3}\left(\frac{h^2 + hk + k^2}{a^2}\right) + \frac{l^2}{c^2} \quad (2)$$

The induced strain ($\varepsilon$) was calculated using the following equation.

$$\varepsilon = \frac{\beta \cos\theta}{4} \quad (3)$$

All the extracted values $2\theta$ at (0 0 2), *d*, FWHM, *D*, c, and $\varepsilon$ are reported in Table 1. The lattice constant (*c*) calculated from the present XRD data is a close match with the standard *c* = 5.205 Å as per JCPDS card No. 005-0664.

**Table 1.** XRD data for samples subjected to various post-growth conditions.

| Post-growth Conditions | $2\theta$ | d (A°) | FWHM | D (nm) | c (A°) | strain ($\varepsilon$) |
|---|---|---|---|---|---|---|
| As grown | 34.405 | 2.604 | 0.110 | 74.969 | 5.209 | 0.026 |
| 350 °C | 34.404 | 2.604 | 0.111 | 74.860 | 5.209 | 0.026 |
| 450 °C | 34.404 | 2.604 | 0.117 | 74.278 | 5.209 | 0.026 |
| LN-15 | 34.407 | 2.604 | 0.114 | 74.592 | 5.208 | 0.026 |
| LN-30 | 34.407 | 2.604 | 0.109 | 75.111 | 5.208 | 0.026 |

From Figure 4, all the samples retain their highly crystalline nature even after the post-growth treatment. It is to note that all the samples are with similar densities as they have grown using the same ammonia concentration. It is interesting to observe that the peak intensity is increased upon increasing the annealing temperature, which signifies the enhanced crystalline quality with respect to the not treated samples. The same is also true for samples treated in liquid nitrogen. Among all the samples, the highest intensity is observed for LN15 sample, which signifies the highest crystallinity in LN15, whereas LN30 has lowered crystalline quality and resembles the sample annealed at 350 °C. The peak intensity is slightly decreased after 30 min in LN but is still above the native sample, meaning that the crystalline quality is not degraded after the post-treatment with either annealing or cryo-cooling. A slight blue-shift of 2θ peak is observed for thermal annealing whereas it is a red-shift for LN treated samples. Increased FWHM and decreased crystallite size are observed for both types of post-growth conditions. These results clearly indicate that the post-growth treatments are modulating the point defects and, in a way, improving the crystalline quality. Overall, LN15 is being the optimum condition among the various post-growth conditions studied to obtain the highest crystallinity. The impact of these various conditions is further validated in terms of their NG device performance.

*3.3. Nanogenerator devices performance*

After duly assessing morphological and structural characterizations of the ZnO-NWs obtained under various conditions, NGs have been fabricated in order to assess their functional piezoelectric properties. Figure 5 (A) presents the schematic test bench and Figure 5 (B) shows the temporal waveform for maximum voltage output obtained at $R_L$ of 50 MΩ of NGs subjected to various post-growth treatments. On Figure 5 (B), the type of post-growth treatment is greatly affecting the NGs output voltage amplitude. The performance of the NGs has been analyzed in terms of their root mean square output voltage ($V_{RMS}$), output current ($I_{RMS}$) and average power ($P_{AV}$), as these are significant values to assess the device performance for practical applications and the results are summarized as follows.

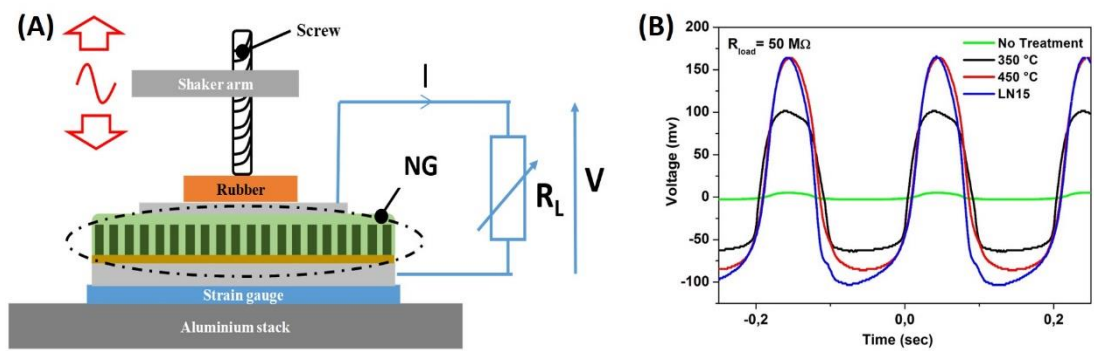

**Figure 5.** (A) Schematic test bench, (B) Temporal waveform of NGs subjected to a 3N compression force at 5Hz, for NWs subjected to different post-growth treatments.

We have applied the various post-growth treatments to the NWs in order to compare their effects on NGs. We have investigated the effect of two distinct approaches, namely thermal annealing in air and cooling in LN, which is a novel approach. Figure 6 (A) and (B) shows the $V_{RMS}$ and $I_{RMS}$, and $P_{AV}$ with respect to varying $R_L$ connected to the NGs, respectively. At a 3N applied periodic force, the post-growth treatment at 450 °C provides better NG performance than the ones treated at 350 °C. At the same 3N force level, the novel post-growth treatment of cryo-cooling in LN for 15 min is resulting in better performance of NGs than both the thermal annealing at 350 °C and 450 °C, which signifies the prominence of the LN cooling. When a 6N force is applied, all the post-growth treatments improve the NG performance but LN15 is still being the best NG among all the studied treatments.

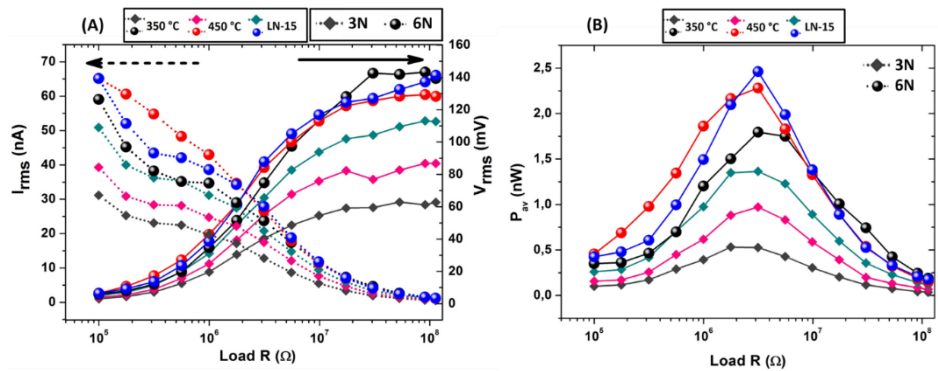

**Figure 6.** Output responses of NGs obtained after various post-growth treatments and subjected to 3N and 6N compression forces at 5Hz; (A) $V_{RMS}$ and $I_{RMS}$ (B) $P_{AV}$.

**Table 2.** Maximum $V_{RMS}$, $I_{RMS}$ and $P_{AV}$ values of NGs prepared under various post-growth conditions.

| Post-growth condition | 3N applied force | | | 6N applied force | | |
|---|---|---|---|---|---|---|
| | $V_{rms}$ (mV) | $I_{rms}$ (nA) | $P_{av}$ (nW) | $V_{rms}$ (mV) | $I_{rms}$ (nA) | $P_{av}$ (nW) |
| As grown | 3.8 | 23.6 | 0.05 | 101 | 49.4 | 1.19 |
| 350 °C | 62 | 31.0 | 0.54 | 141 | 58.7 | 1.94 |
| 450 °C | 86 | 38.8 | 0.97 | 129 | 71.1 | 2.35 |
| LN-15 | 114 | 50.7 | 1.38 | 144 | 63.9 | 2.51 |
| LN-30 | 3.2 | 24.2 | 0.05 | N/A | N/A | N/A |

Table 2 summarizes the maximum NG output performances for all the tested devices prepared under various post-growth conditions. The maximum values wherein reported for $I_{RMS}$, $V_{RMS}$, and $P_{AV}$ are observed at load values 100 kΩ, 50 MΩ, and 2.5 MΩ respectively. Clearly, an increment in the device performances can be seen for post-growth treatments, thermal annealing at both the temperatures and cryo-cooling for 15 min. The NG performance is increased with the increased applied force. Moreover, an increment on performances can be observed in Figure 7 for these post-growth conditions. Thermal annealing in air at 450 °C is showing a superior performance than at 350 °C. It was reported that thermal annealing modulates the free charge carrier density, in agreement to which the present study finds the enhanced NG performance. A further enhancement in NG performance is observed for cryo-cooling in LN for 15 min duration. However, an extended duration of cryo-cooling (i.e.30 min) reduced the device performance reaching back to the untreated condition. Therefore, LN-15 condition can be understood as an optimum to obtain the highest performance of NGs. It is hypothesized that the longer duration of LN cooling may form the moisture deep inside the NWs and develop a passivating layer, thereby limiting the electrical performance. Annealing at mild temperatures may be able to resolve this issue. In accordance with the crystallinity analysis (Table 1), these post-growth treatments modulate the point defects and are thereby affecting the free charge carriers inside the NWs. Therefore, it can be concluded that among all the post-growth conditions investigated, cryo-cooling in LN for 15 min is a promising, as well as a feasible green solution avoiding the high-temperature treatments to obtain NGs with high performance. Of course, this preliminary study should be followed by an extensive study of the cryo-cooling treatment duration, and its effect on the NWs defects.

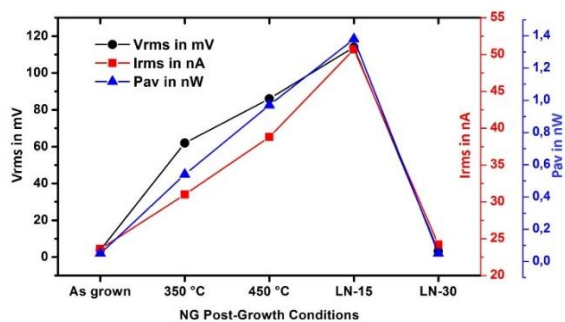

**Figure 7.** Performance of NGs subjected to a 3N compression force at 5Hz, with ZnO-NWs subjected to various post-growth treatments.

## 4. Conclusions

In summary, we have demonstrated strategies for enhancing the performance of piezoelectric NGs with ZnO-NWs arrays on rigid Si substrates. First, ZnO-NWs with high density were grown, using low-cost, and scalable bottom-up process on gold-coated surfaces.

In order to further enhance the NG performances, thermal annealing in air and cryo-cooling in liquid nitrogen have been applied as post-growth treatments and their eventual effect is investigated. It is observed that none of the post-growth treatments affected the morphology of the nanostructures. Clearly, significant improvements in crystalline quality have been observed upon thermal annealing at 350 °C and 450 °C and cryo-cooling for 15 min, whereas crystalline quality is lowered upon cryo-cooling for a longer duration of 30 min but still remains higher than the not treated sample. Corresponding changes observed in crystallite size are ascribed to the modulation of point defects which thereby affects the free charge carrier concentration in ZnO-NWs.

Interestingly, very similar improvements in the performance of the NGs corresponding to the improved crystalline quality upon post-growth treatment are observed. Cryo-cooling for 15 min has shown to lead to very promising performance of NG and to be competitive to the thermal annealing at 450 °C. Therefore, this novel approach as a post-treatment offers an alternative solution where thermal annealing has a limitation, for example for flexible polymer substrate-based NGs [26]. We justify that the improved performances of NGs are due to enhancement of the piezoelectric potential via free-carrier passivation in the hydrothermal ZnO-NWs subjected to post-growth treatments. Therefore, these approaches provide a simple and viable solution to enhance the NG performances for realizing self-powered electronics.

**Funding:** EnSO project has been accepted for funding within the Electronic Components and Systems For European Leadership Joint Undertaking in collaboration with the European Union's H2020 Framework Programme (H2020/2014-2020) and National Authorities [Grant agreement No. 692482]. This work was also supported by the French National Research Agency [Grant agreement ANR-14-CE08-0010-01]; and the Region Centre [Grant agreements 2014-00091629 and 2016-00108356].

**Conflicts of Interest:** The authors declare no conflict of interest.

# Strategies to Enhance ZnO Nanogenerator Performance

# via Thermal-Annealing and Cryo-Cooling

Guylaine Poulin-Vittrant, Kevin Nadaud, Chandrakanth Reddy Chandraiahgari and Daniel Alquier

GREMAN UMR 7347, CNRS, Université de Tours, INSA-CVL

16 rue Pierre et Marie Curie, BP 7155, 37071 TOURS cedex 2, FRANCE

Correspondence: guylaine.poulin-vittrant@univ-tours.fr

# Supporting information

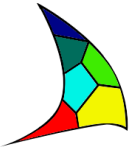
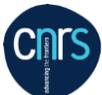
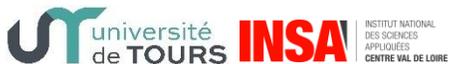

1. Effect of Varying Concentration of Ammonia

Varying amounts of Ammonia (0-40 mM) were introduced in the growth solution in order to study the effect of ammonia.

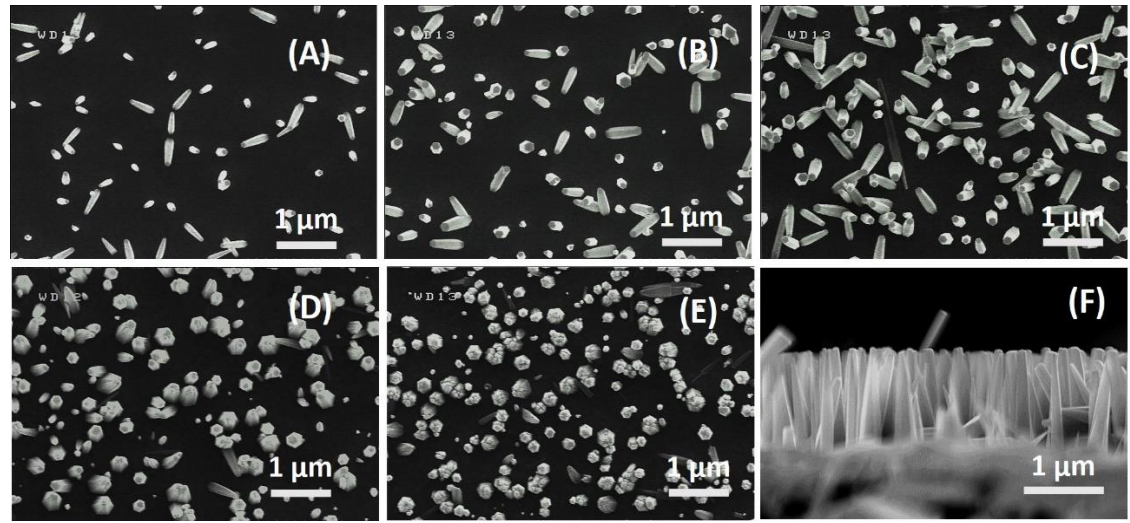

SEM images of ZnO-NWs grown with different concentrations of ammonia: (A) 0 mM, (B) 10 mM, (C) 20 mM, (D) 30 mM, (E) 40 mM, and (F) cross section of NWs grown with 30mM ammonia

1. Effect of Varying Concentration of Ammonia

Varying amounts of Ammonia (0-40 mM) were introduced in the growth solution in order to study the effect of ammonia.

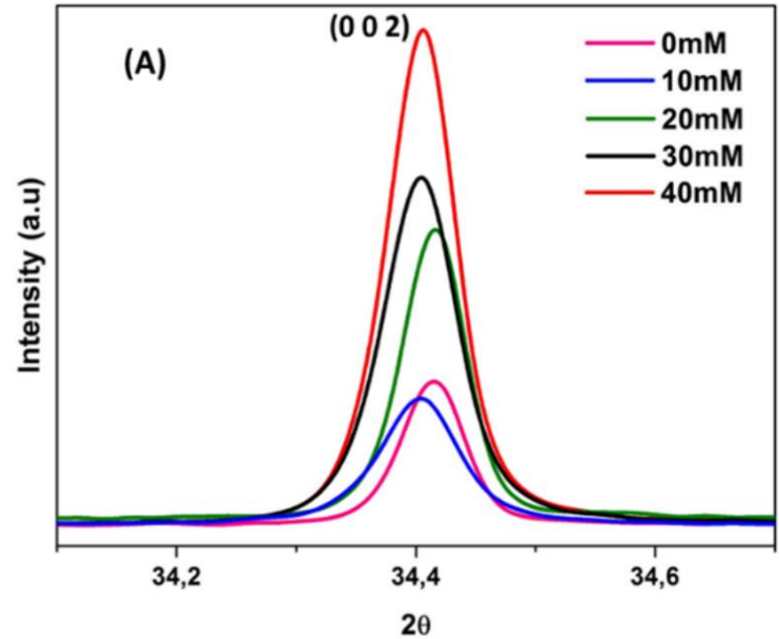

XRD patterns of ZnO-NWs produced by varying ammonia concentration

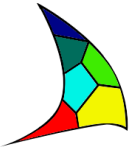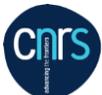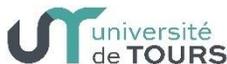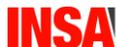

# Nanowires on Ti/Au-Si : Growth using 30 mM Ammonia and No Annealing

$V_{RMS}$ : Root Mean Square voltage
$I_{RMS}$ : Root Mean Square current
$P_{AV}$ : Average power

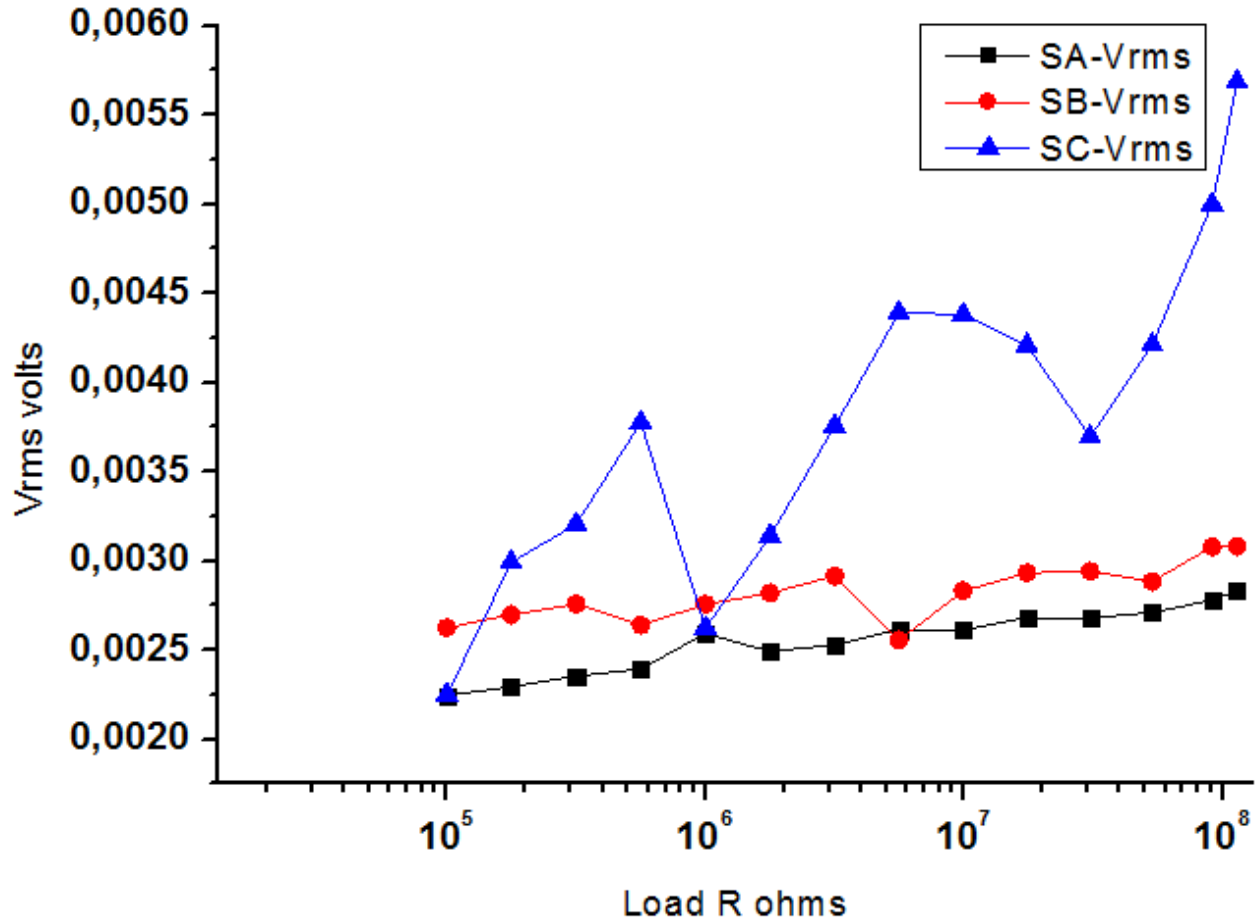
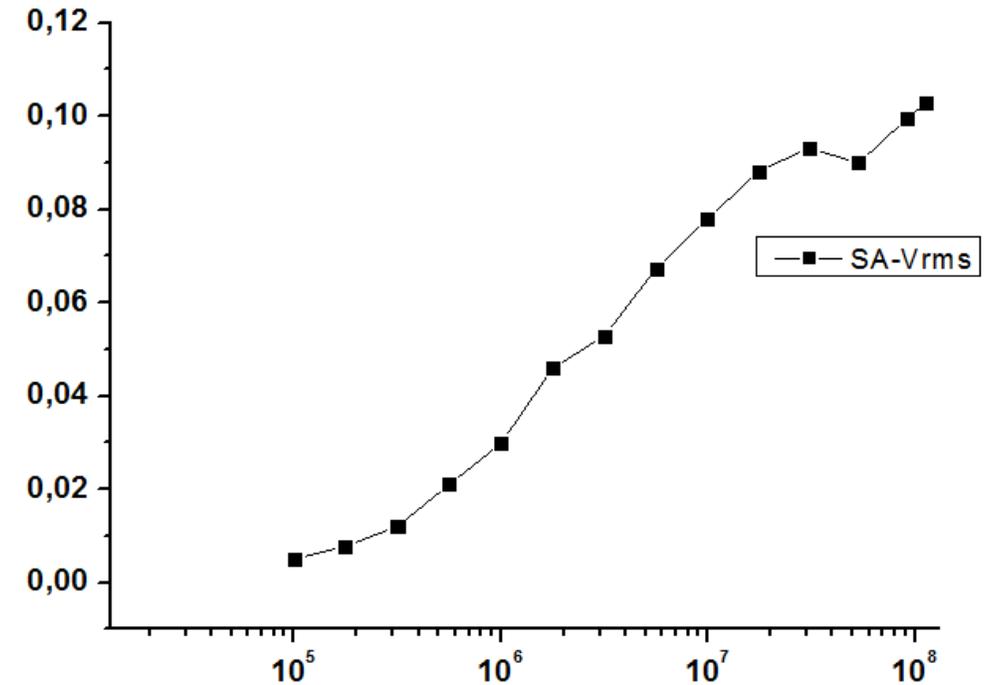

$V_{RMS}$ delivered by the NGs subjected to a compression force at 5Hz: under 3N (left) and 6N (right)

# Nanowires on Ti/Au-Si : Growth using 30 mM Ammonia and No Annealing

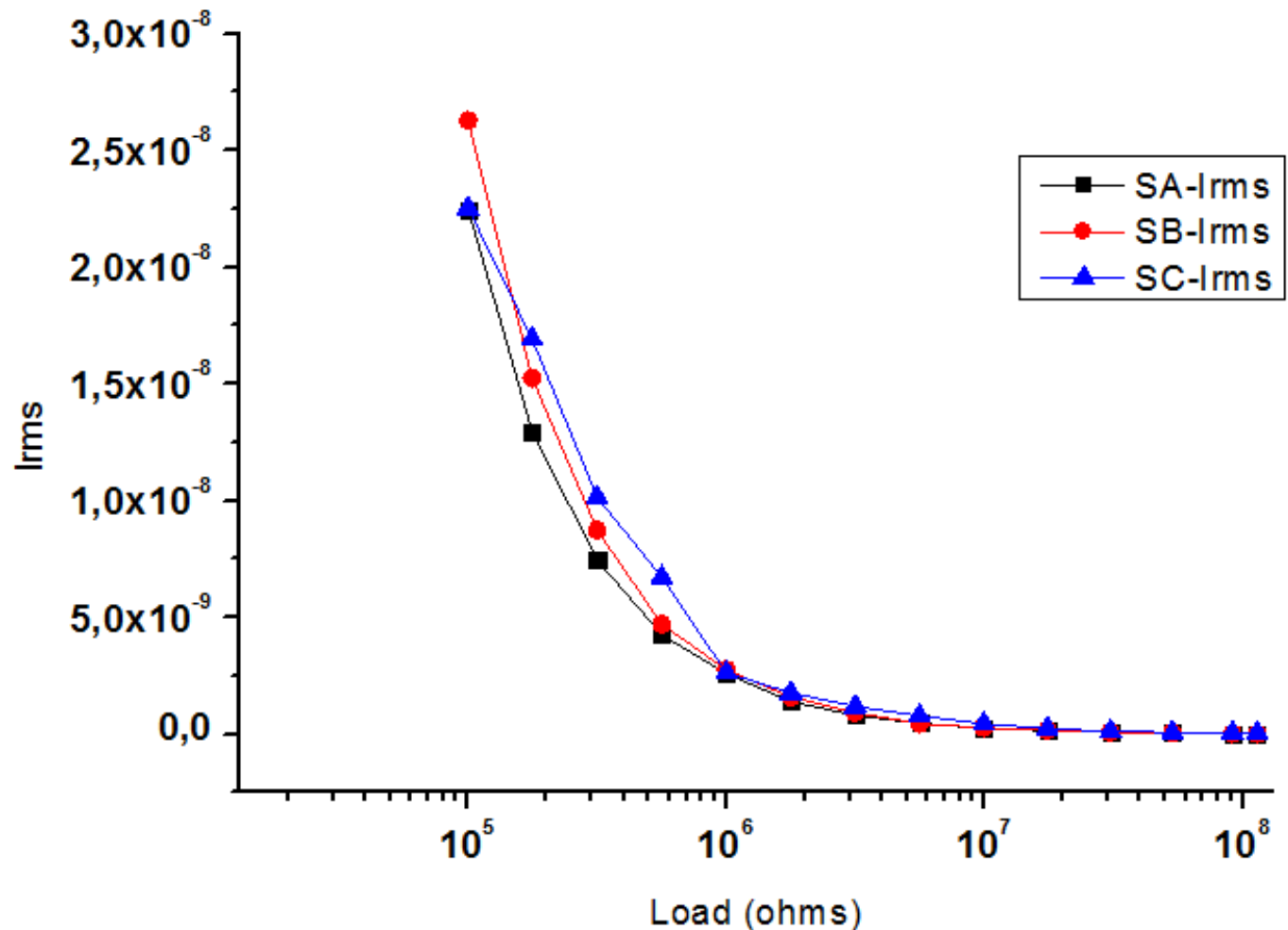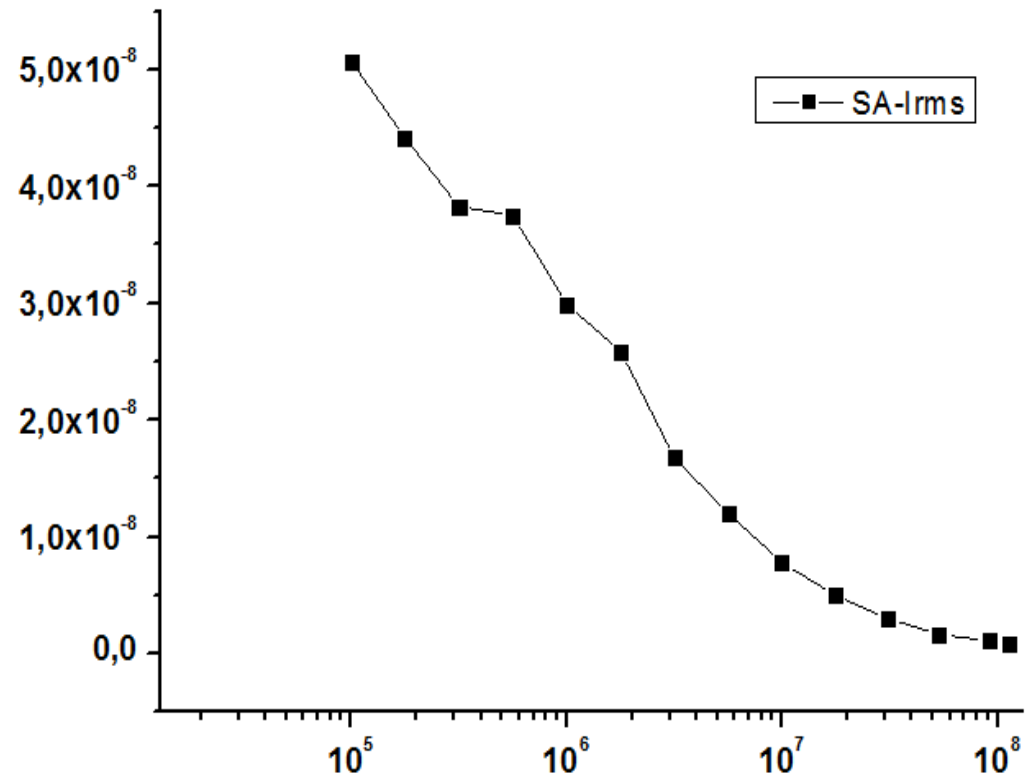

$I_{RMS}$ delivered by the NGs subjected to a compression force at 5Hz: under 3N (left) and 6N (right)

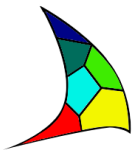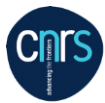

# Nanowires on Ti/Au-Si : Growth using 30 mM Ammonia and No Annealing

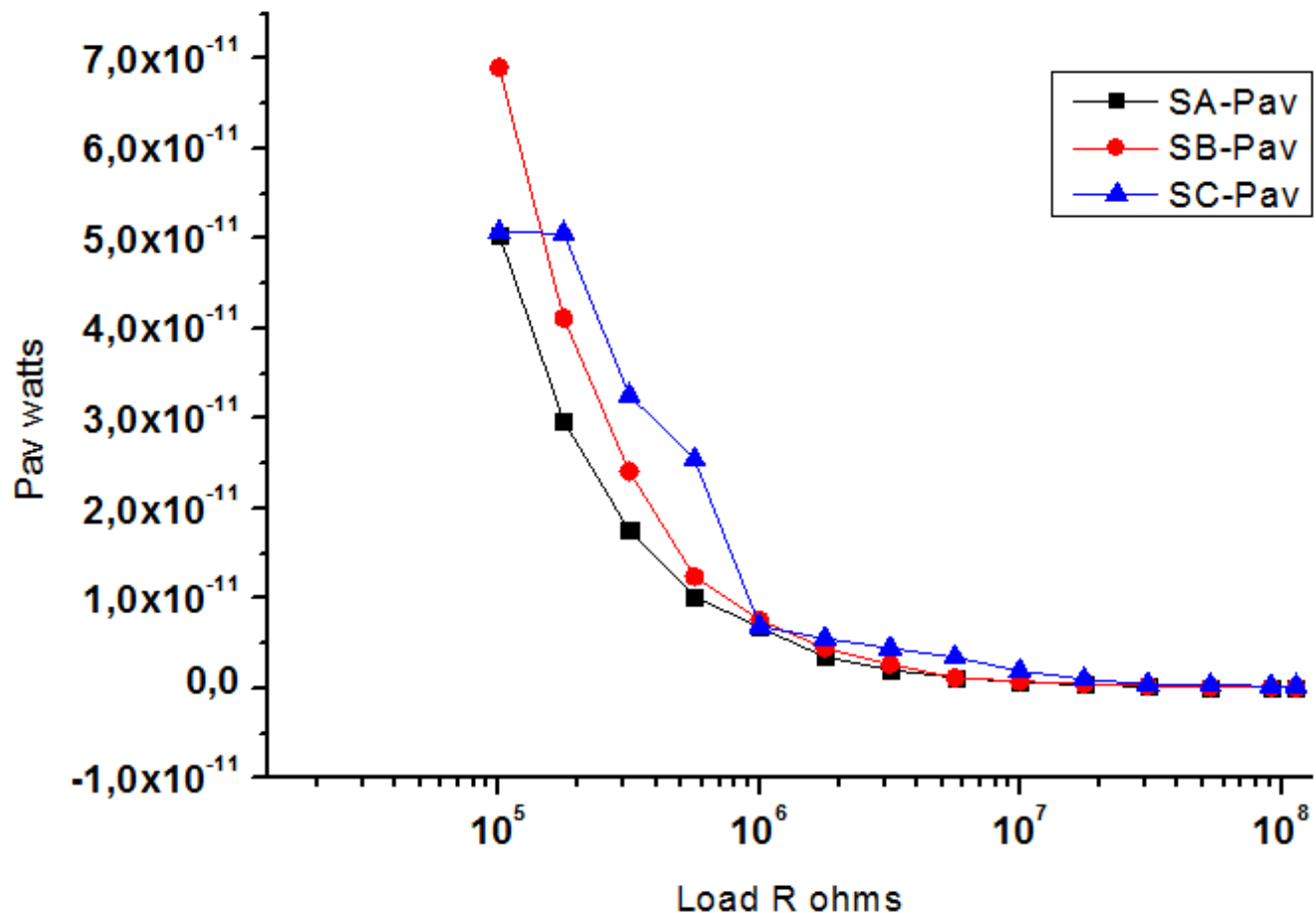
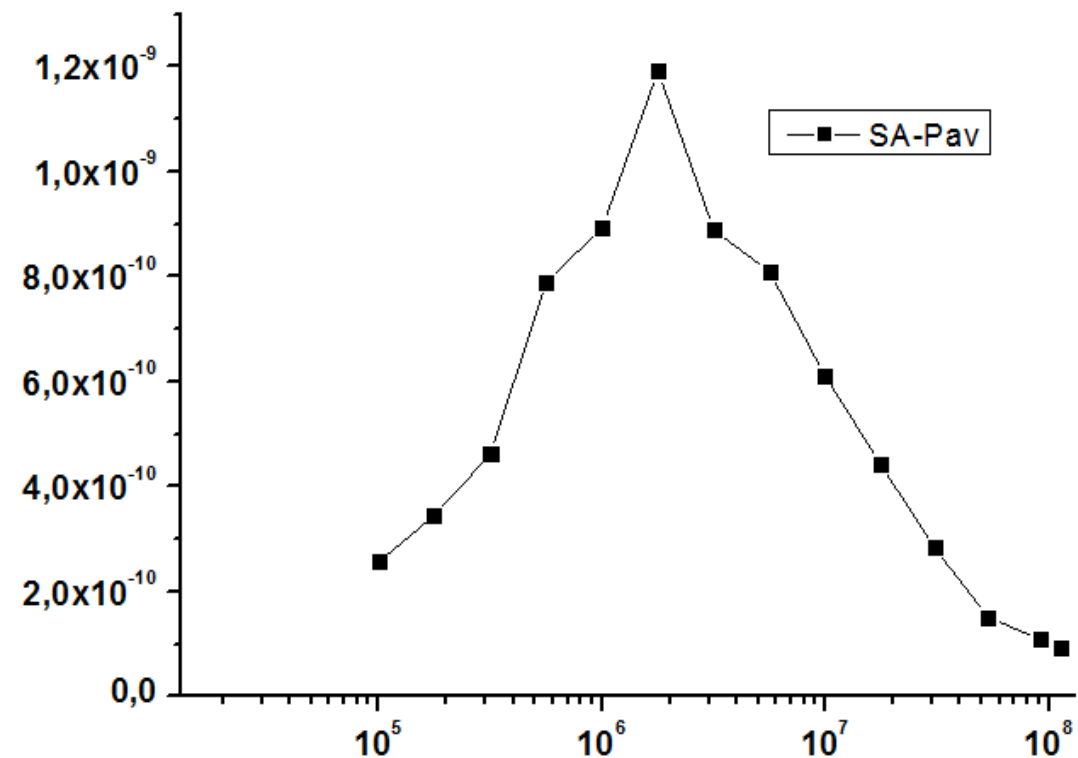

$P_{AV}$ delivered by the NGs subjected to a compression force at 5Hz: under 3N (left) and 6N (right)

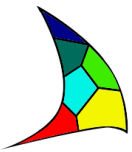
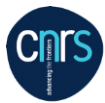

# Nanowires on Ti/Au-Si : Growth using 30 mM Ammonia and 350°C Annealing for 90 min

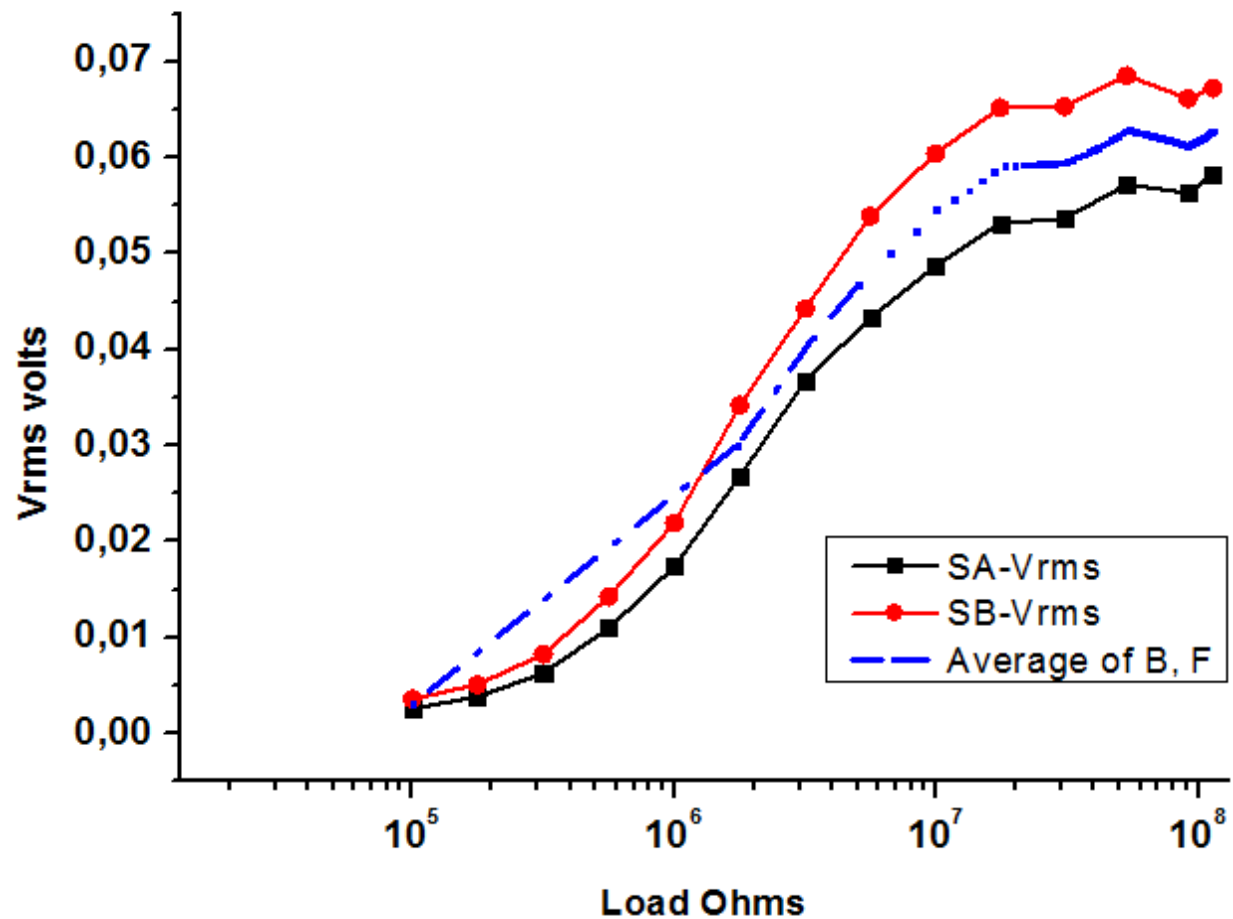
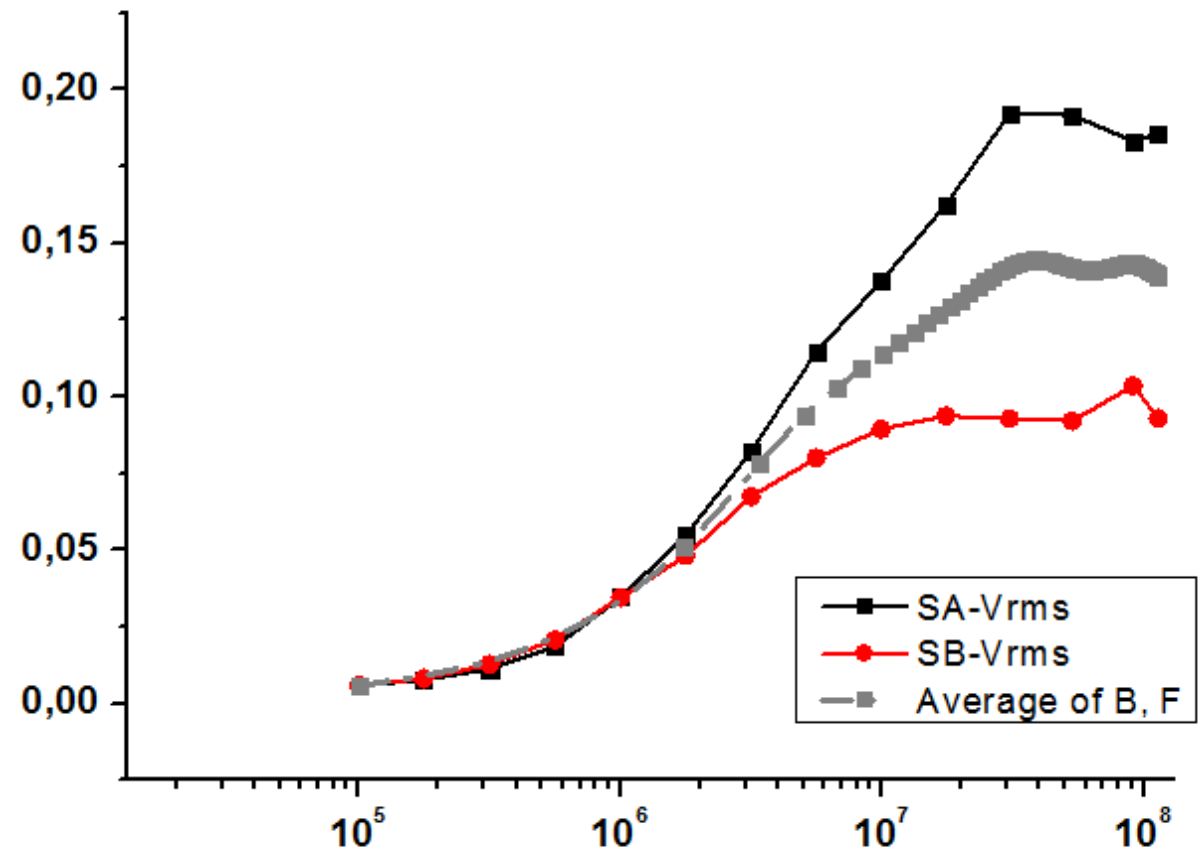

$V_{RMS}$ delivered by the NGs subjected to a compression force at 5Hz: under 3N (left) and 6N (right)

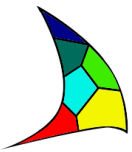
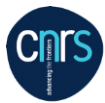

# Nanowires on Ti/Au-Si : Growth using 30 mM Ammonia and 350°C Annealing for 90 min

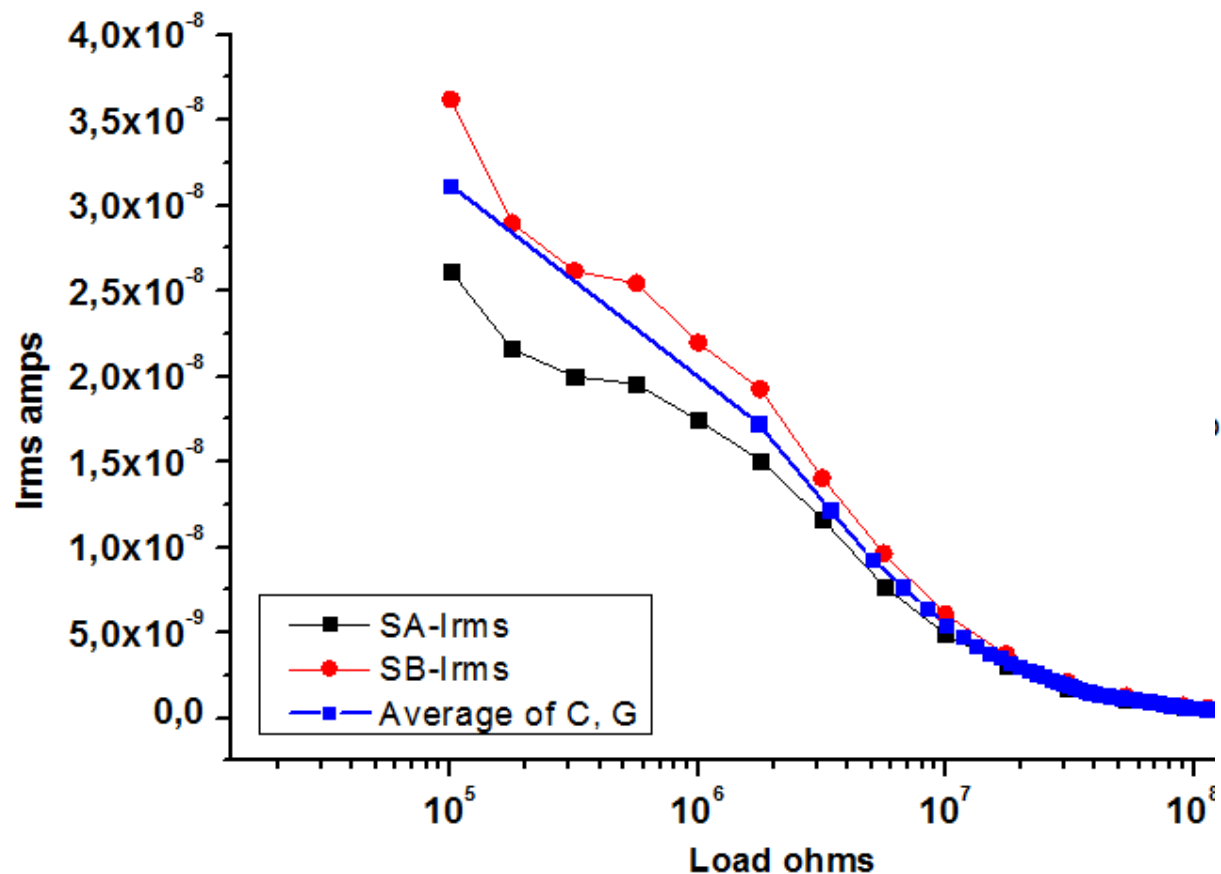
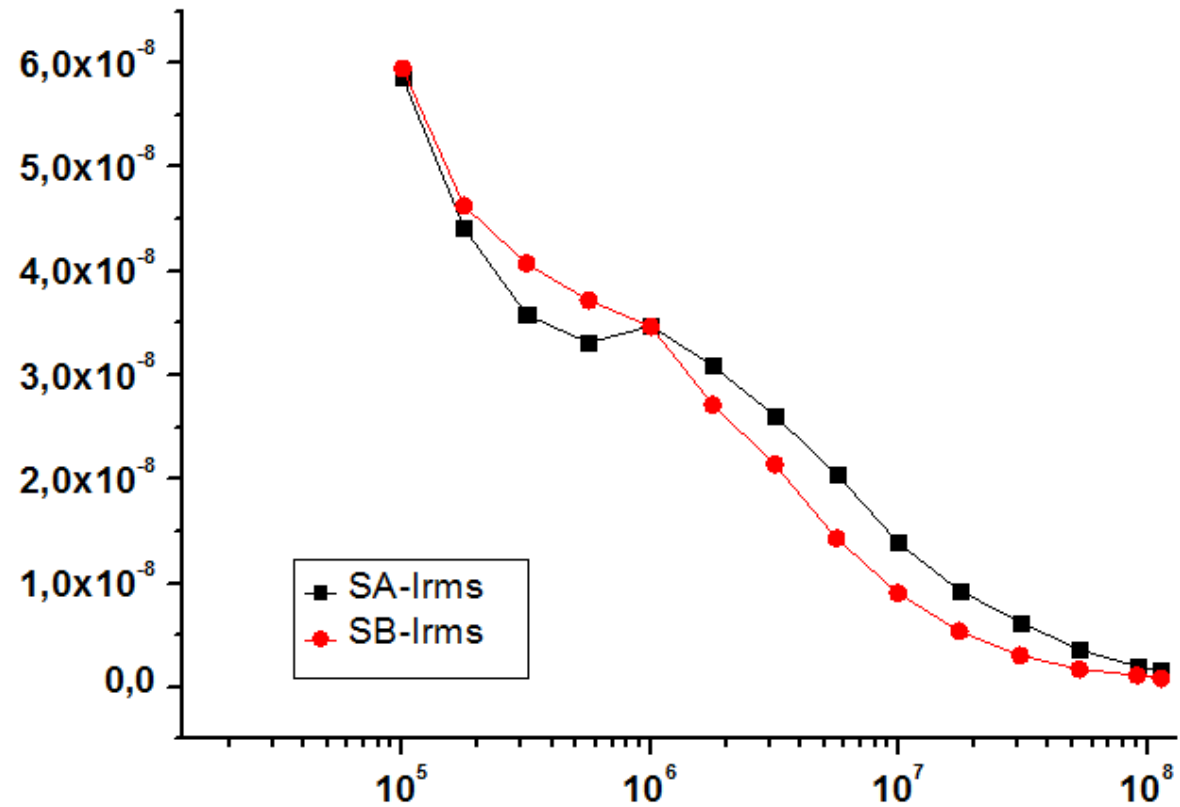

$I_{RMS}$ delivered by the NGs subjected to a compression force at 5Hz: under 3N (left) and 6N (right)

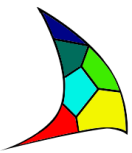
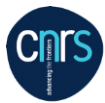

# Nanowires on Ti/Au-Si : Growth using 30 mM Ammonia and 350°C Annealing for 90 min

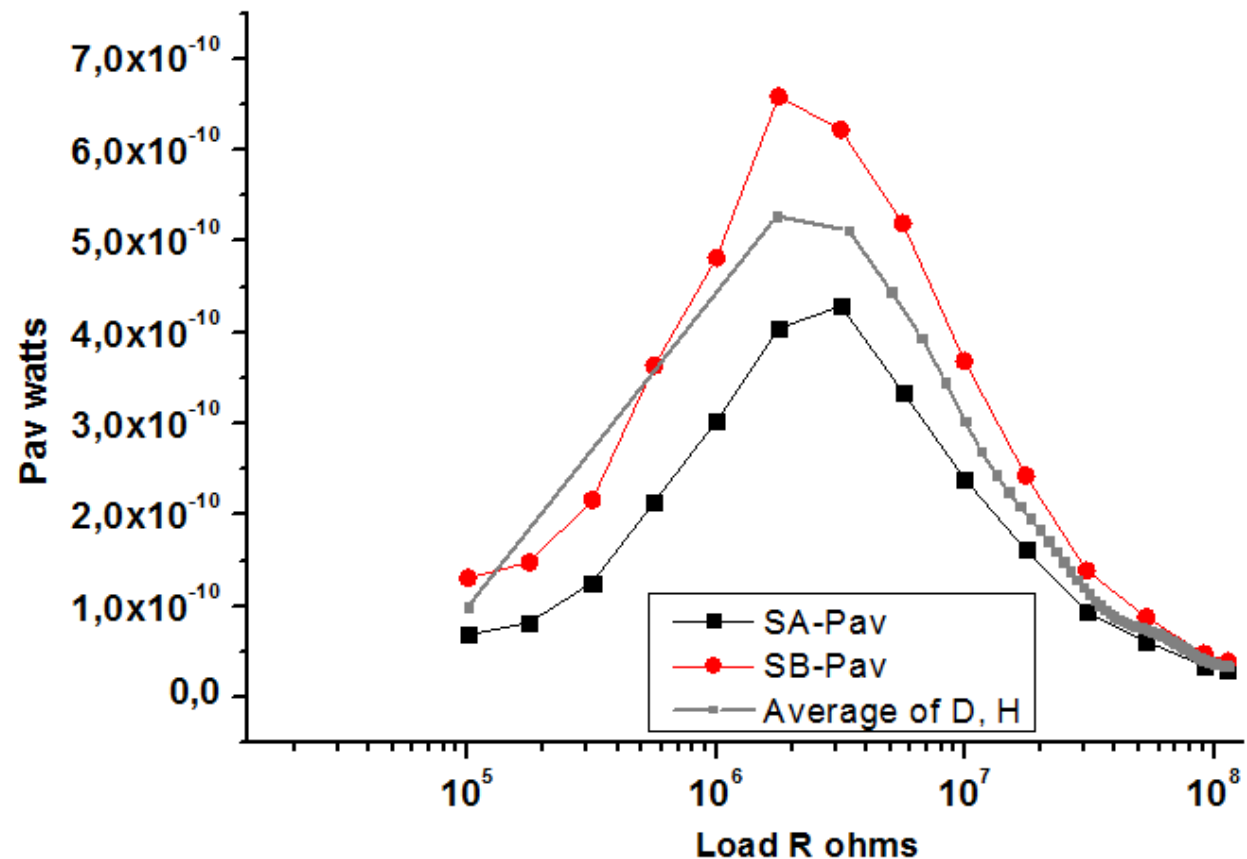
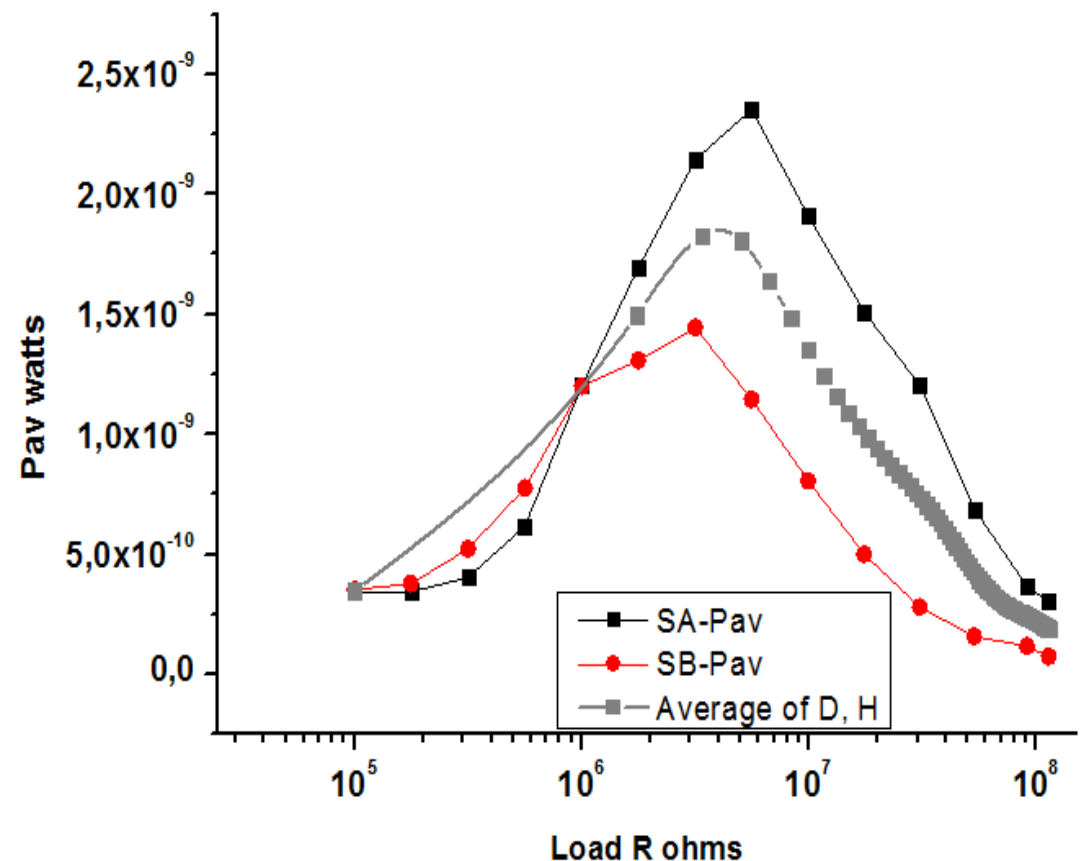

$P_{AV}$ delivered by the NGs subjected to a compression force at 5Hz: under 3N (left) and 6N (right)

# Nanowires on Ti/Au-Si : Growth using 30 mM Ammonia and 450 °C Annealing for 30 min

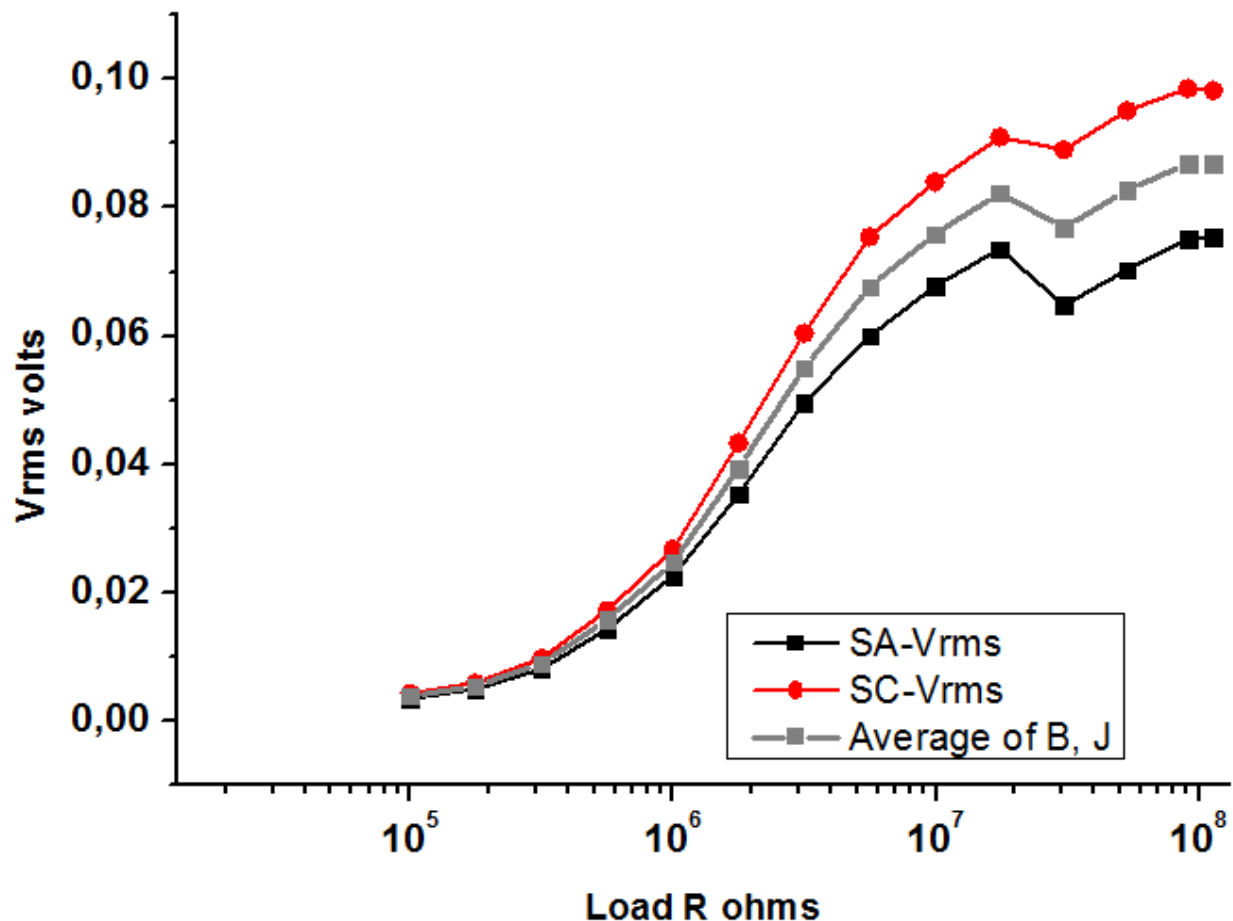
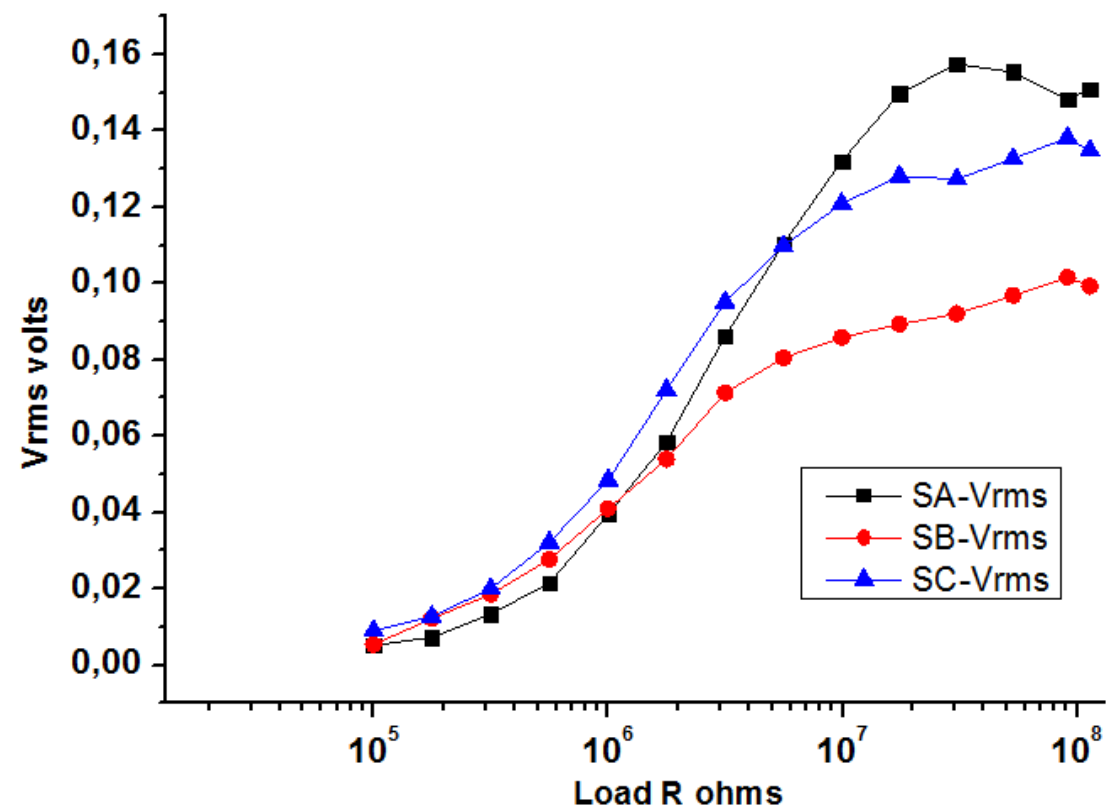

$V_{RMS}$ delivered by the NGs subjected to a compression force at 5Hz: under 3N (left) and 6N (right)

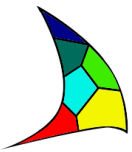
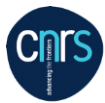

# Nanowires on Ti/Au-Si : Growth using 30 mM Ammonia and 450 °C Annealing for 30 min

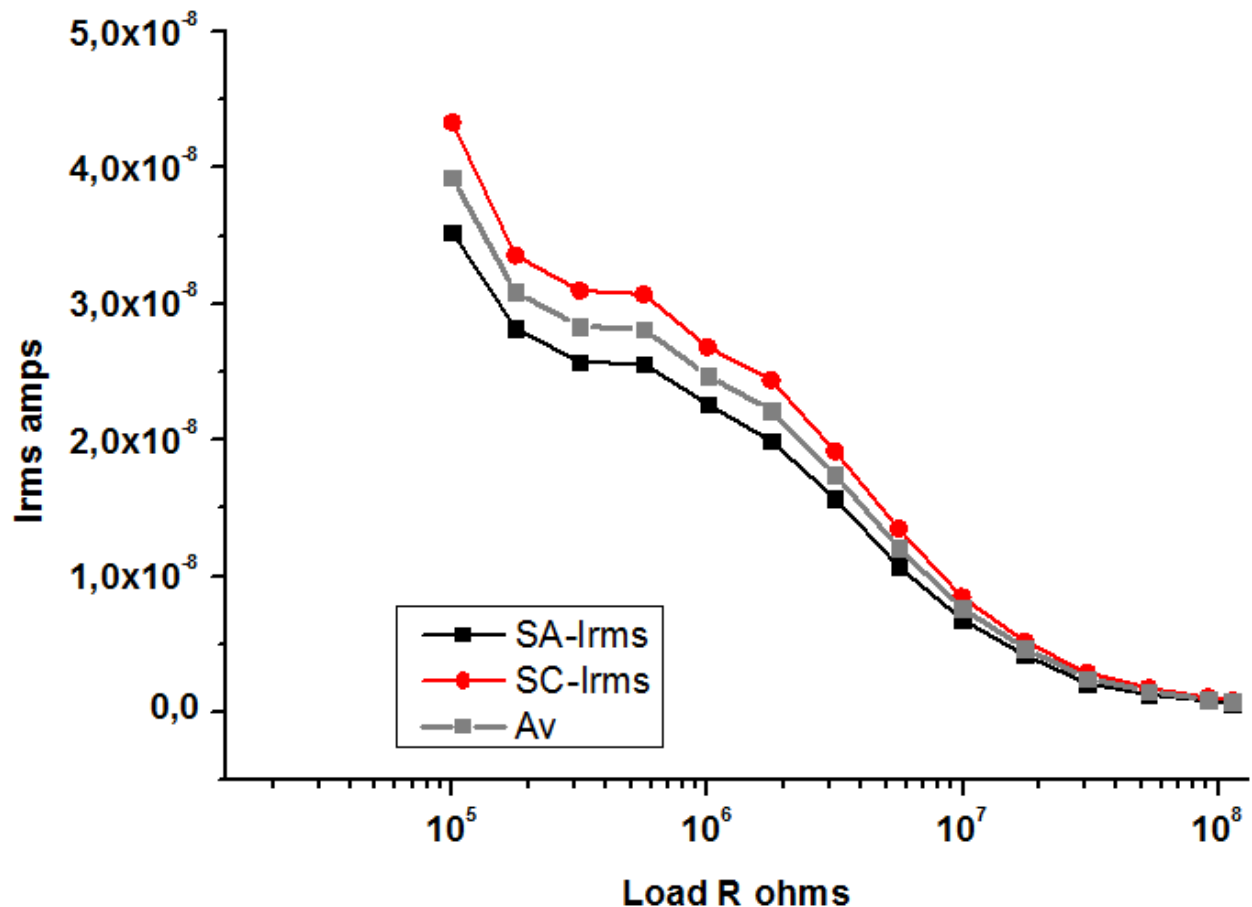 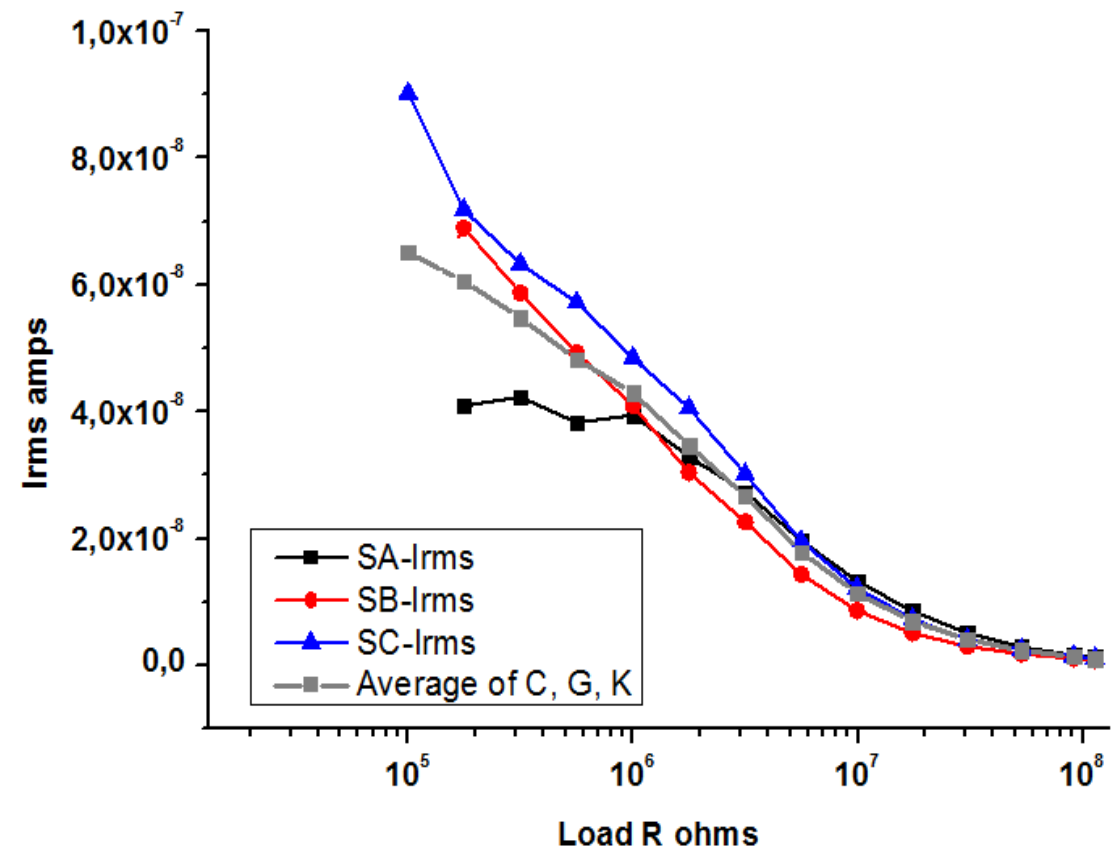

$I_{RMS}$ delivered by the NGs subjected to a compression force at 5Hz: under 3N (left) and 6N (right)

# Nanowires on Ti/Au-Si : Growth using 30 mM ammonia and 450 °C Annealing for 30 min

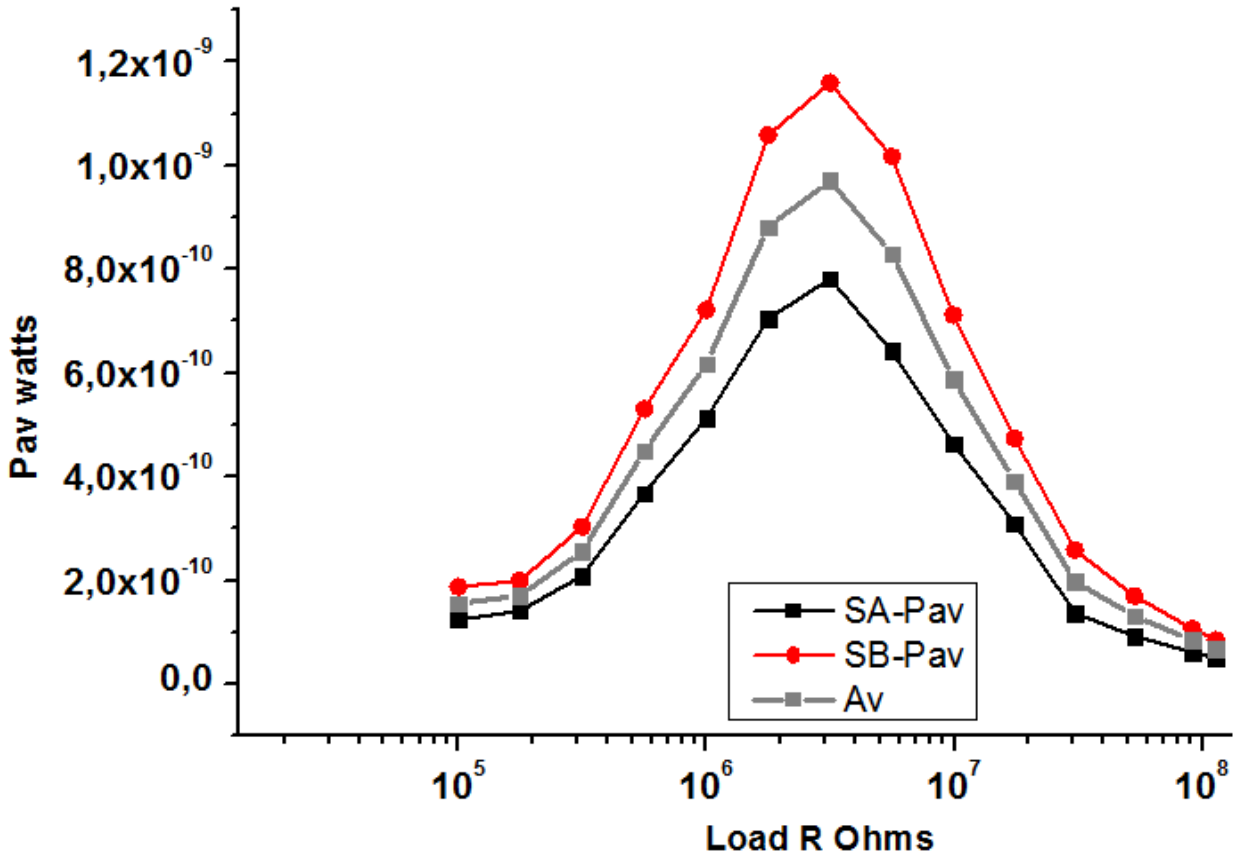 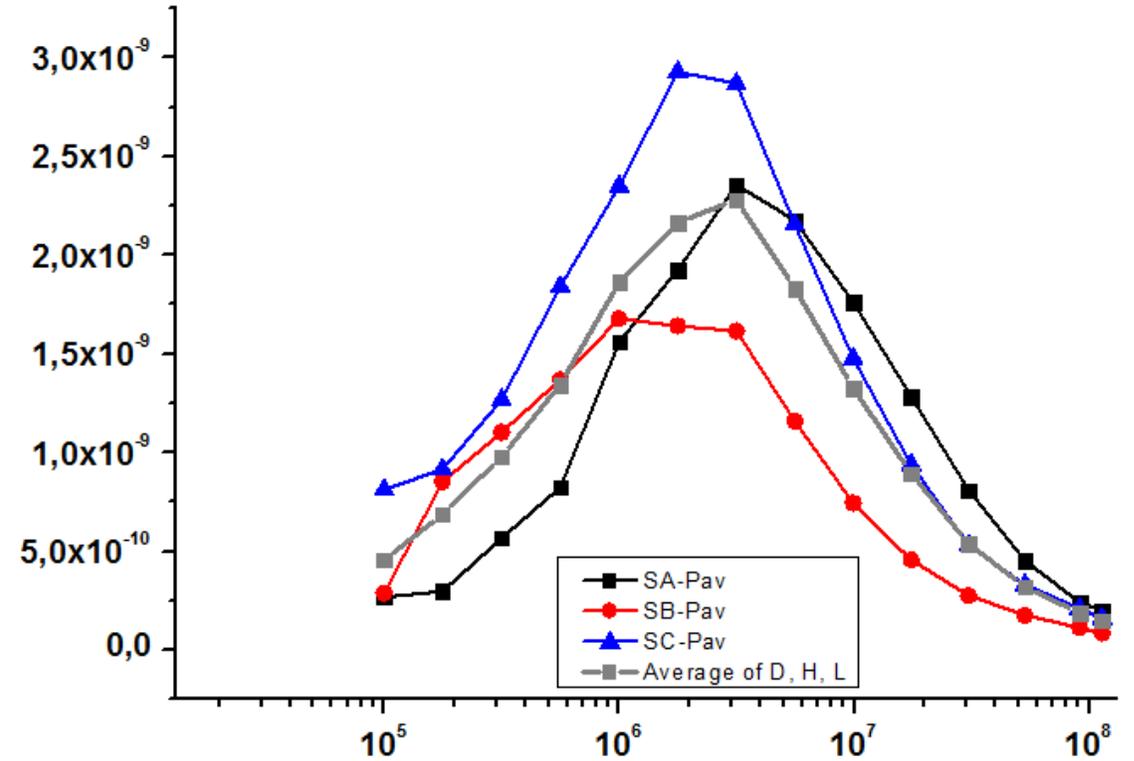

$P_{AV}$ delivered by the NGs subjected to a compression force at 5Hz: under 3N (left) and 6N (right)

# Nanowires on Ti/Au-Si : Growth using 40 mM Ammonia and Annealing 350 °C for 90 min

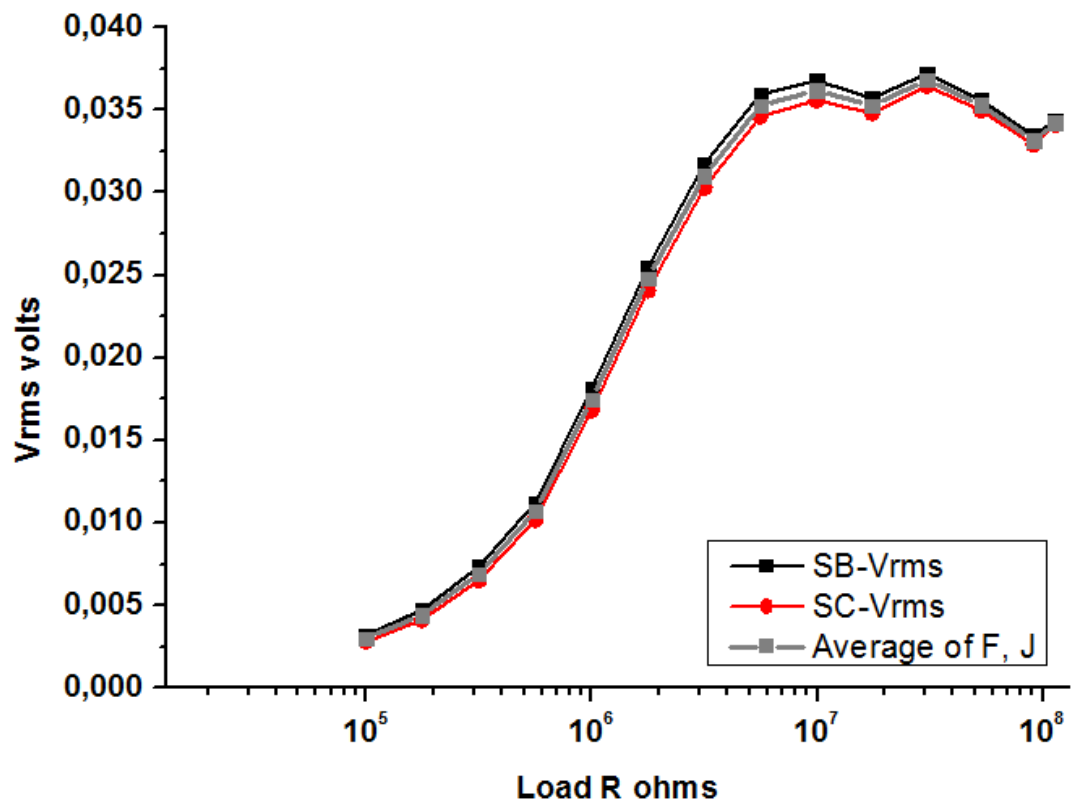

$V_{RMS}$ delivered by the NGs subjected to a 3N compression force at 5Hz

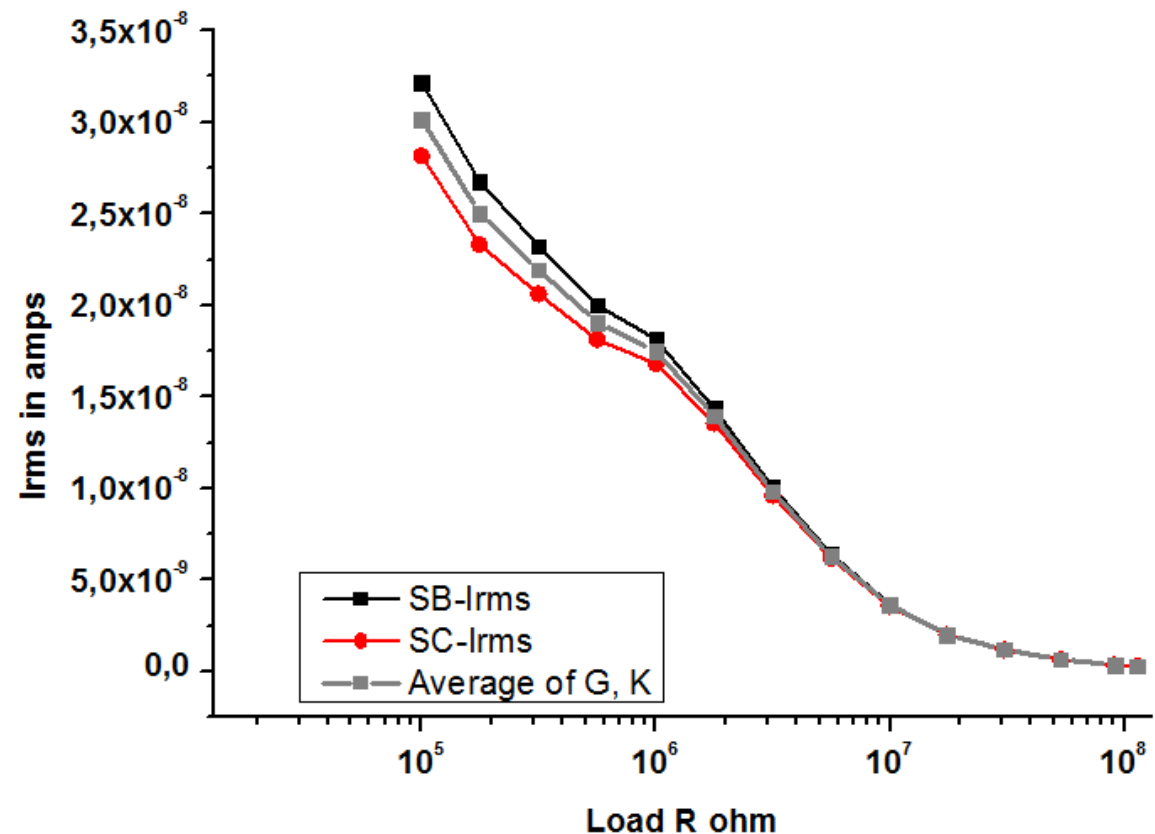

$I_{RMS}$ delivered by the NGs subjected to a 3N compression force at 5Hz

# Nanowires on Ti/Au-Si : Growth using 40 mM Ammonia and Annealing 350 °C for 90 min

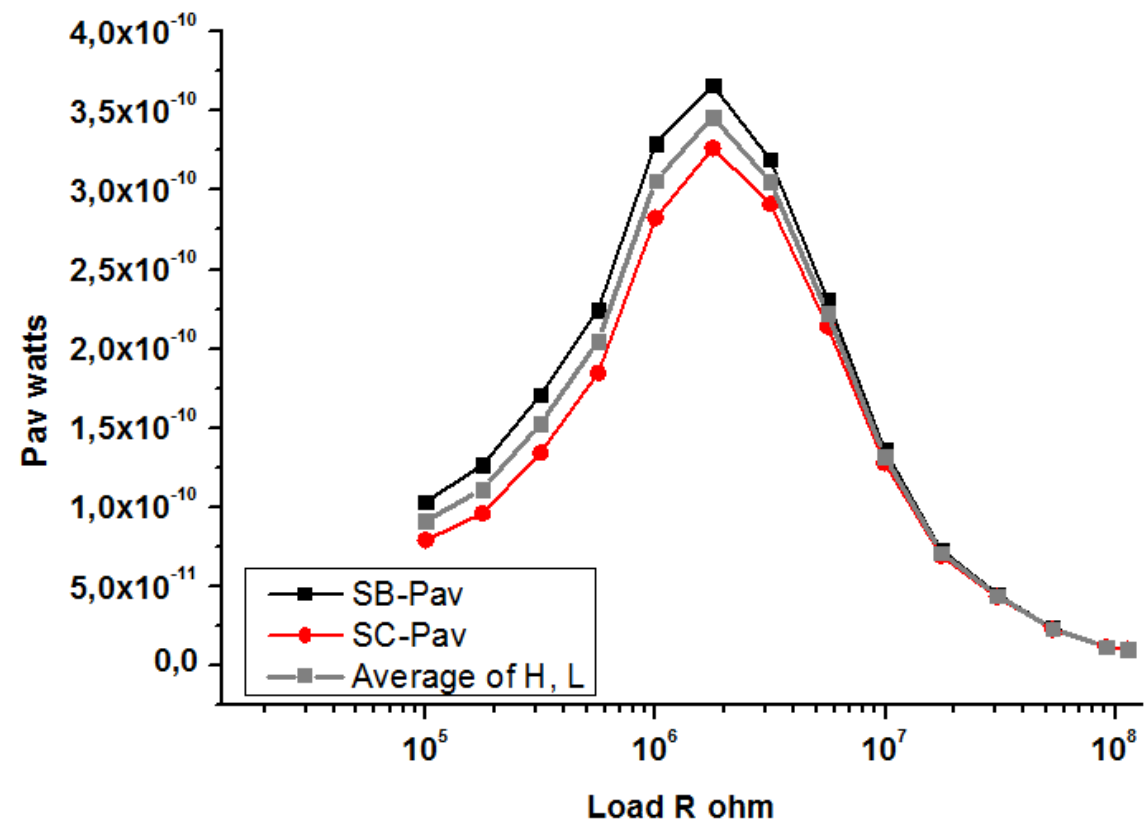

$P_{AV}$ delivered by the NGs subjected to a 3N compression force at 5Hz

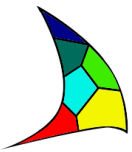
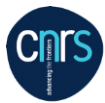

# Nanowires on Ti/Au-Si : Growth using 20 mM Ammonia and Annealing 350 °C for 90 min

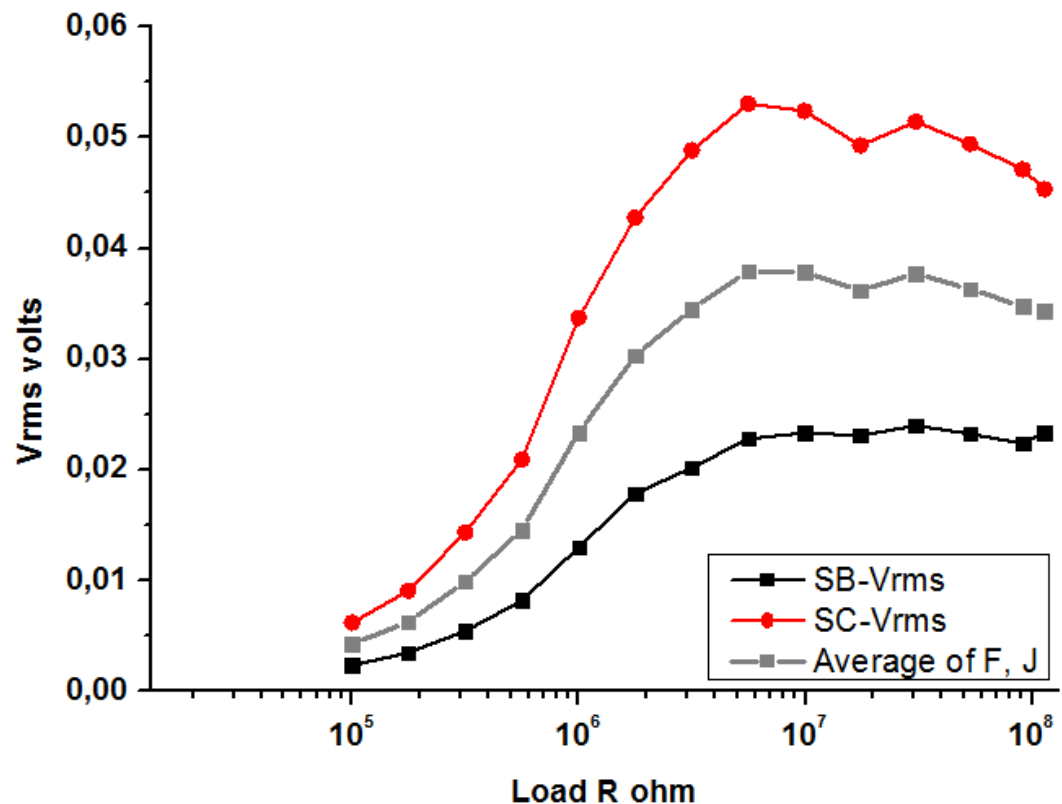

$V_{RMS}$ delivered by the NGs subjected to a 3N compression force at 5Hz

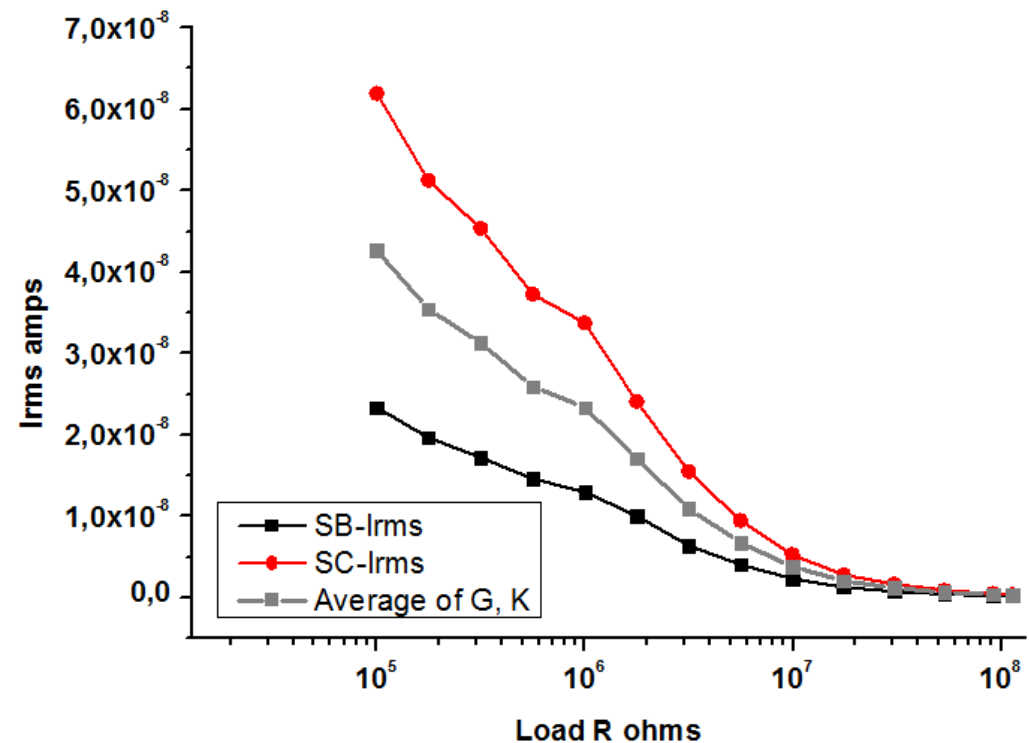

$I_{RMS}$ delivered by the NGs subjected to a 3N compression force at 5Hz

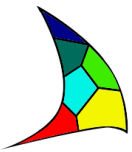
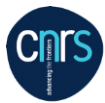

# Nanowires on Ti/Au-Si : Growth using 20 mM Ammonia and Annealing 350 °C for 90 min

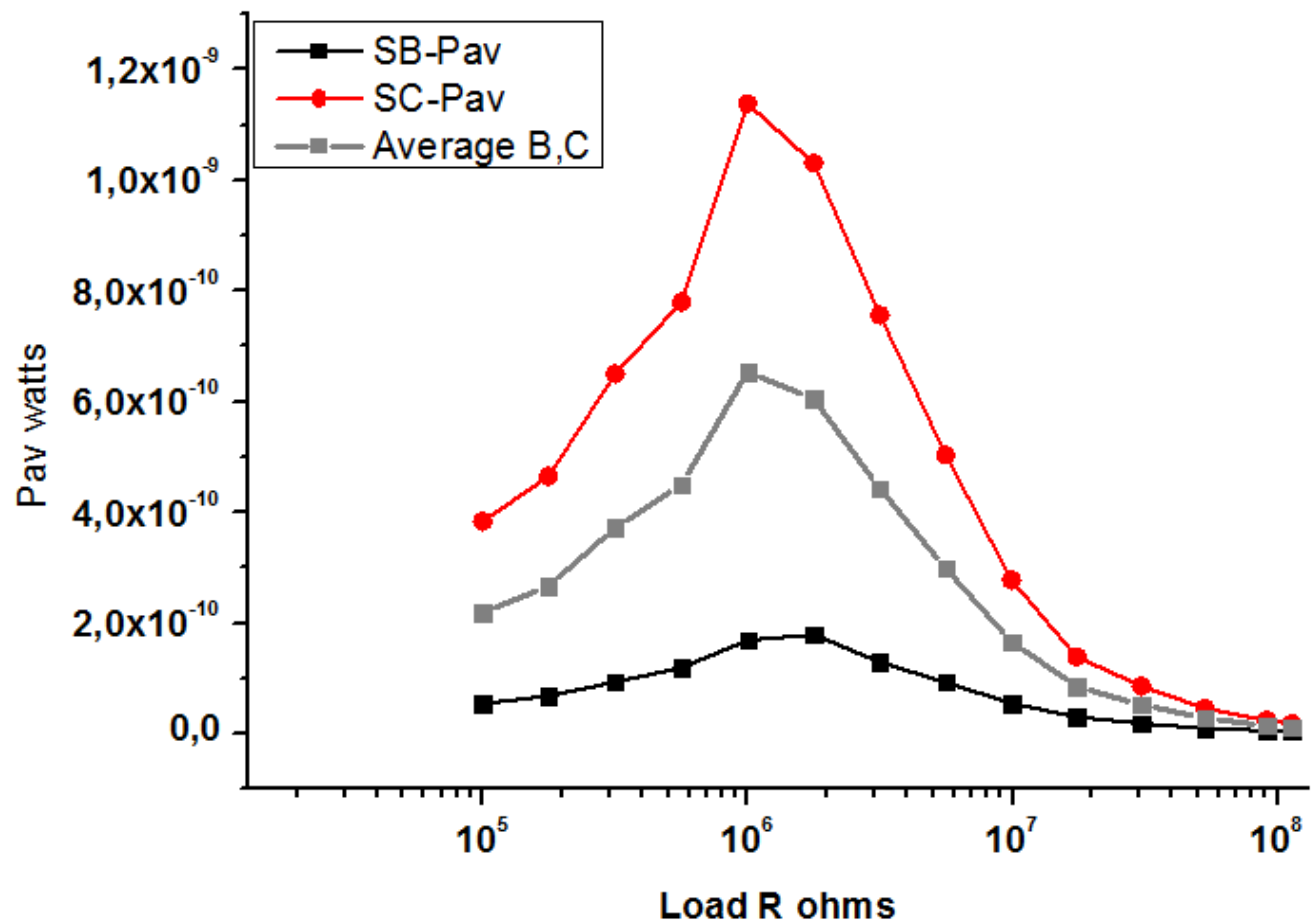

$P_{AV}$ delivered by the NGs subjected to a 3N compression force at 5Hz

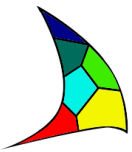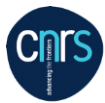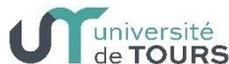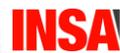

# Nanowires on Ti/Au-Si
# Growth using 0 to 40 mM Ammonia and 350 °C Annealing for 90 min

| Sample Name | at 3N Force | | |
| --- | --- | --- | --- |
| | $V_{RMS}$ (V) | $I_{RMS}$ (A) | $P_{mean}$ (W) |
| Si_TiAu_0mM_growth_350°C annealing | 0.0022 | 1.87e-8 | 0.03e-9 |
| Si_TiAu_10mM_growth_350°C annealing | 0.0025 | 2.12e-8 | 0.07e-9 |
| Si_TiAu_20mM_growth_350°C annealing | 0.022 | 2.34e-8 | 0.18e-9 |
| Si_TiAu_30mM_growth_350°C annealing | 0.062 | 3.1e-8 | 0.54e-8 |
| Si_TiAu_40mM_growth_350°C annealing | 0.034 | 2.95e-8 | 0.34e-9 |

Maximum $V_{RMS}$, $I_{RMS}$ and $P_{AV}$ values of NGs prepared with varying ammonia concentration, and annealing in air for 90 min

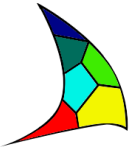
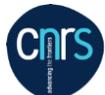
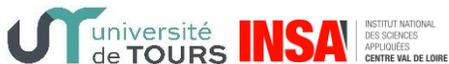

## 2. Effect of post-treatments:
Thermal annealing and Immersion in Liquid Nitrogen.

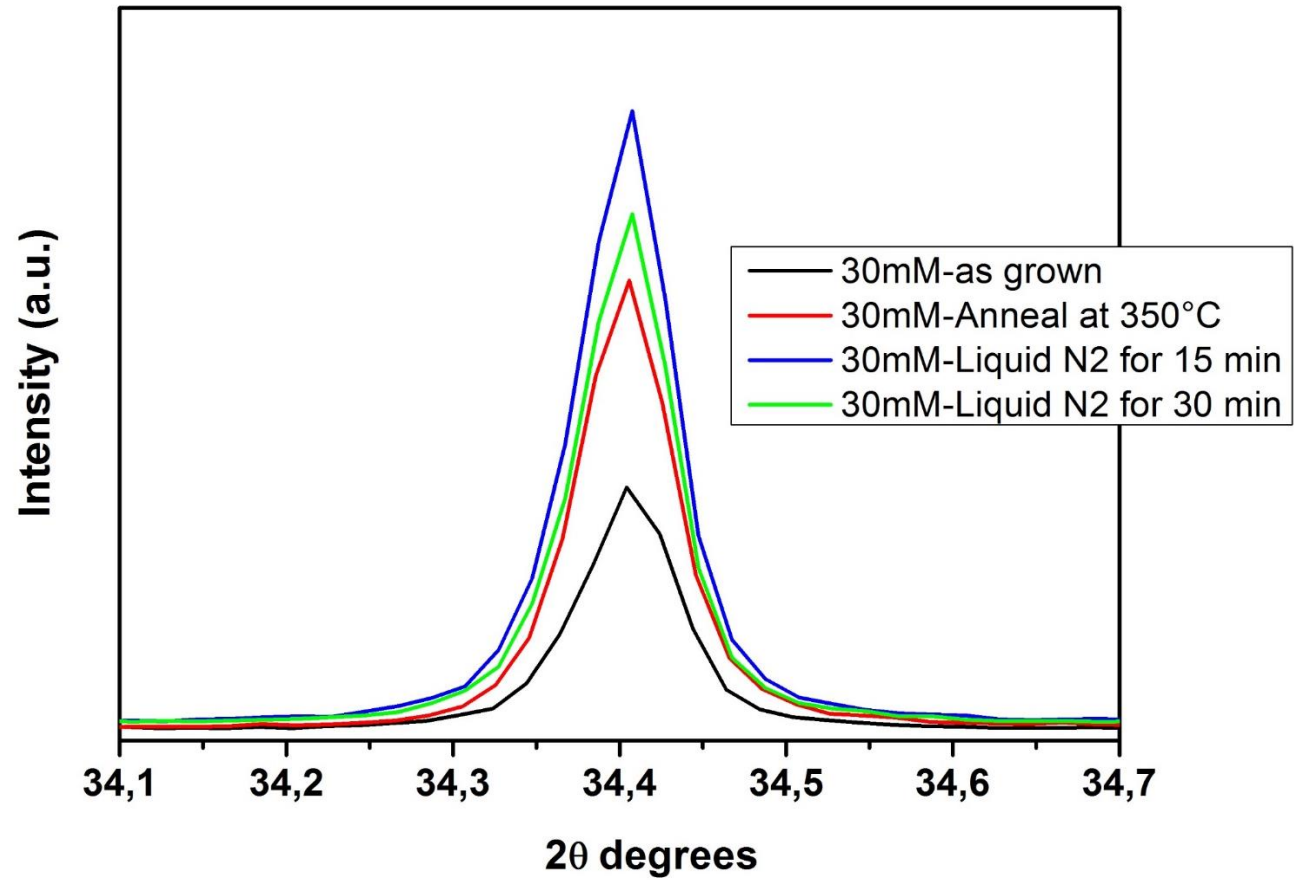

XRD patterns of ZnO-NWs grown with 30 mM ammonia concentration and subjected to various post-treatments

# 2. Effect of post-treatments:
## Thermal annealing and Immersion in Liquid Nitrogen

| Sample Name | 2Θ° | Interplanar Spacing (d) A° | FWHM | Crystallite Size (D) nm |
|---|---|---|---|---|
| 0mM_Growth_350C Anneal | 34.4241 | 2.60317 | 0.10057 | 82,69608101 |
| 10mM_Growth_350°C Anneal | 34.4078 | 2.60436 | 0.10064 | 82,63492142 |
| 20mM_Growth_350°C Anneal | 34.4075 | 2.60438 | 0.10658 | 78,02938404 |
| 30mM_Growth_350°C Anneal | 34.4058 | 2.60451 | 0.11093 | 74,96920177 |
| 40mM_Growth_350°C Anneal | 34.4045 | 2.60461 | 0.12129 | 68,56545752 |
| 30mM_Growth_No Anneal | 34.4043 | 2.60462 | 0.11109 | 74,86092221 |
| 40mM_Growth_No Anneal | 34.4045 | 2.60461 | 0.12129 | 68,56545752 |
| 30mM_Growth_Liquid N2_15 min | 34.4075 | 2.60439 | 0.11149 | 74,59298368 |
| 30mM_Growth_Liquid N2_30 min | 34.4075 | 2.60438 | 0.11072 | 75,11173908 |

XRD data for ZnO-NWs produced with varying ammonia concentrations and subjected to various post-treatments

# Nanowires on Ti/Au-Si : Growth using 30 mM Ammonia and Immersion in Liquid N$_2$ for 15 min

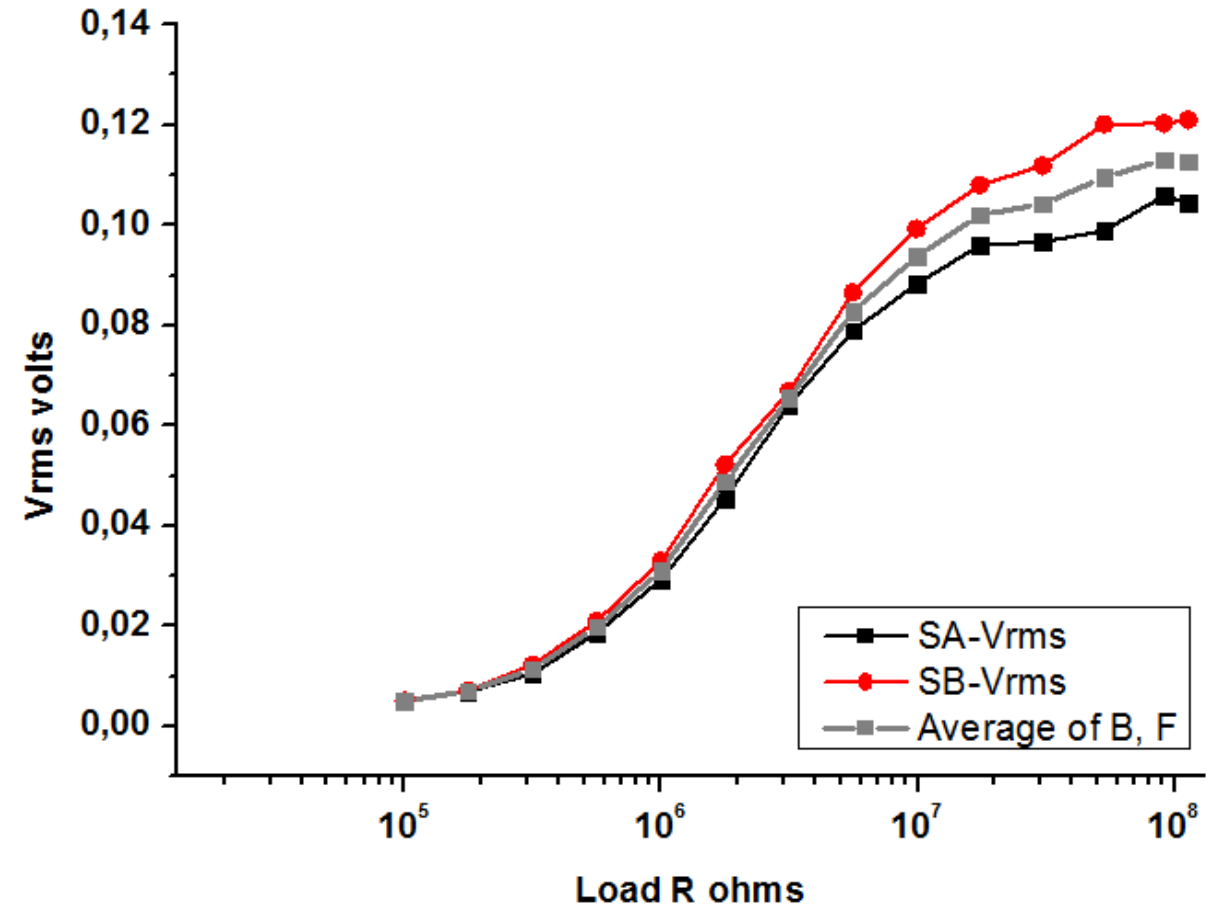
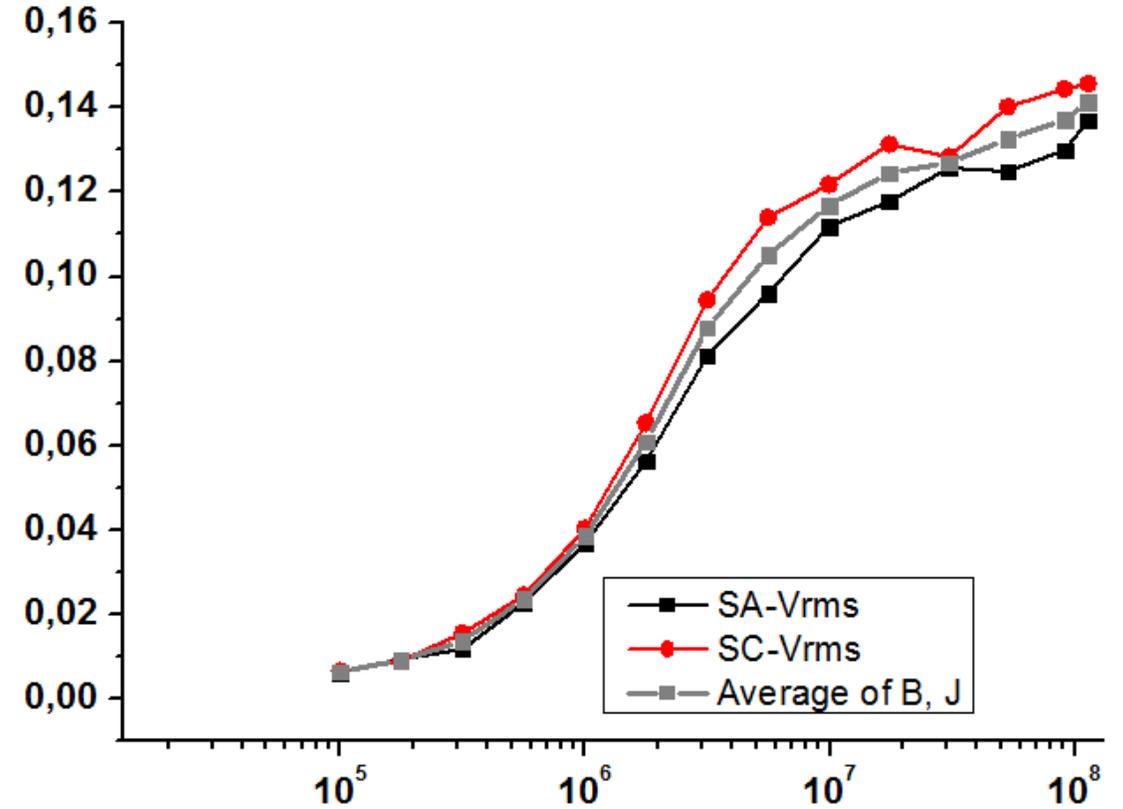

V$_{RMS}$ delivered by the NGs subjected to a compression force at 5Hz: under 3N (left) and 6N (right)

# Nanowires on Ti/Au-Si : Growth using 30 mM Ammonia and Immersion in Liquid $N_2$ for 15 min

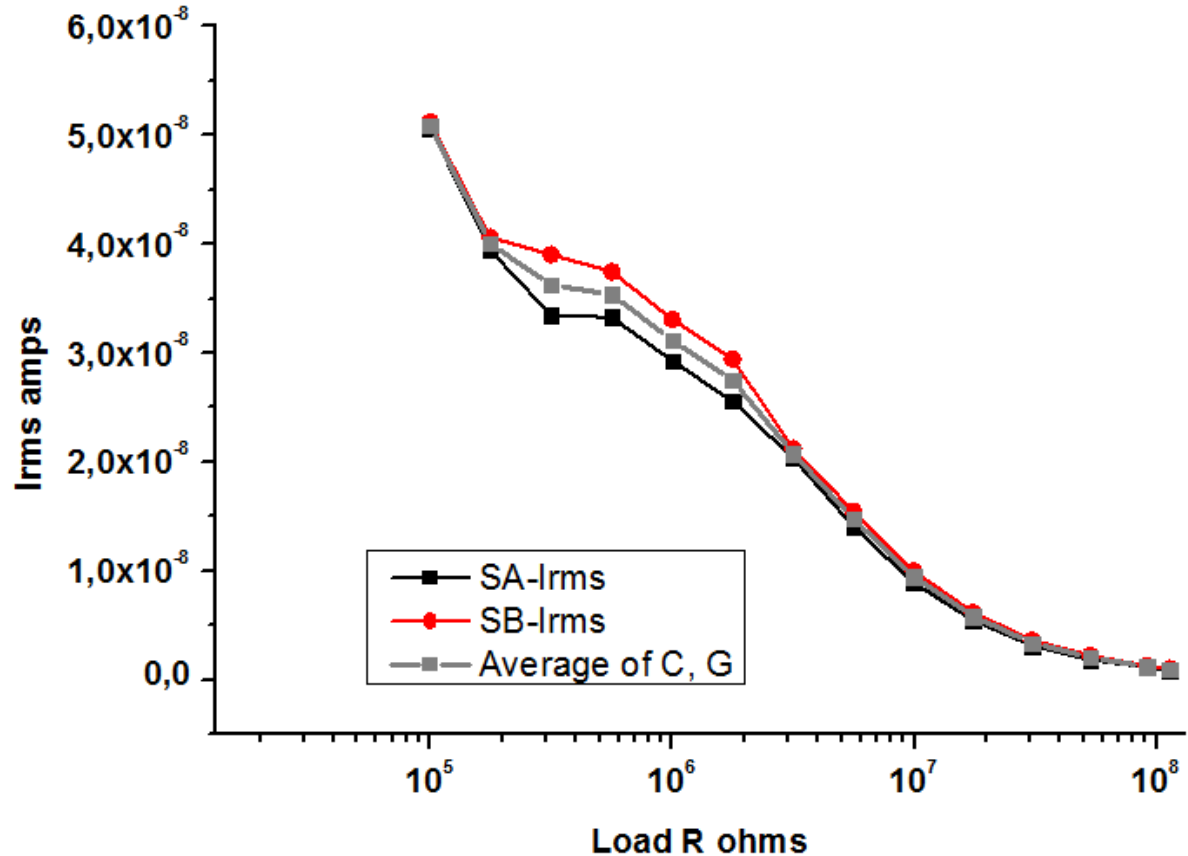
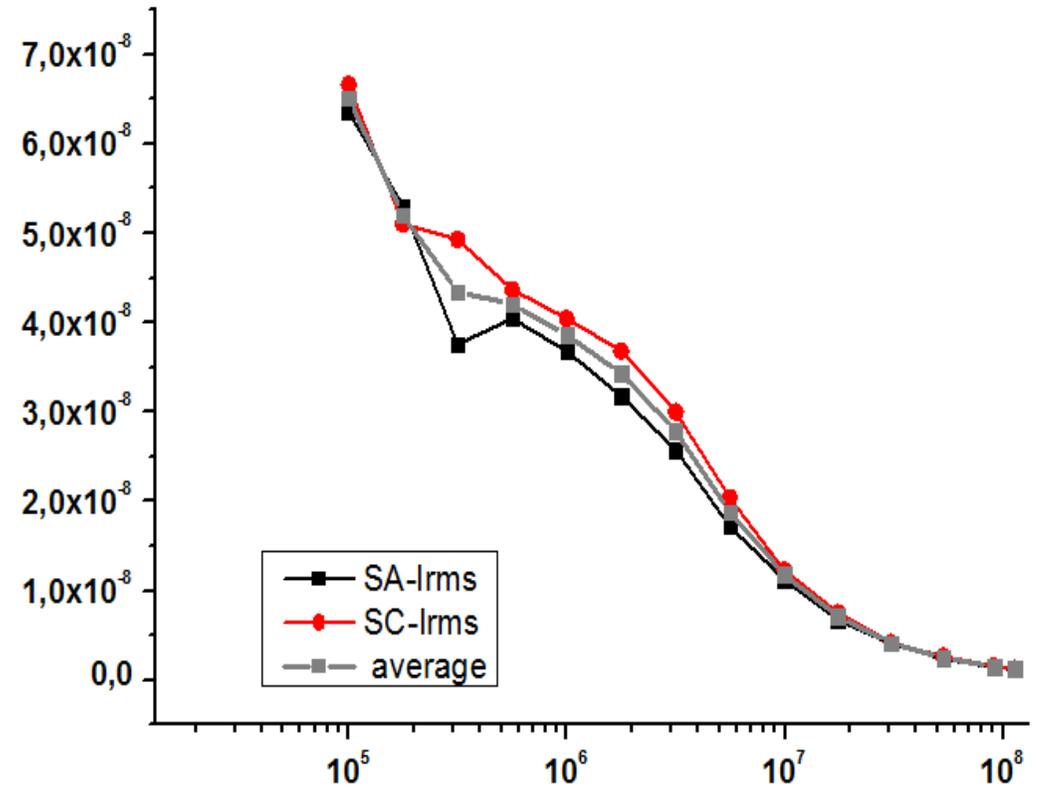

$I_{RMS}$ delivered by the NGs subjected to a compression force at 5Hz: under 3N (left) and 6N (right)

# Nanowires on Ti/Au-Si : Growth using 30 mM Ammonia and Immersion in Liquid N$_2$ for 15 min

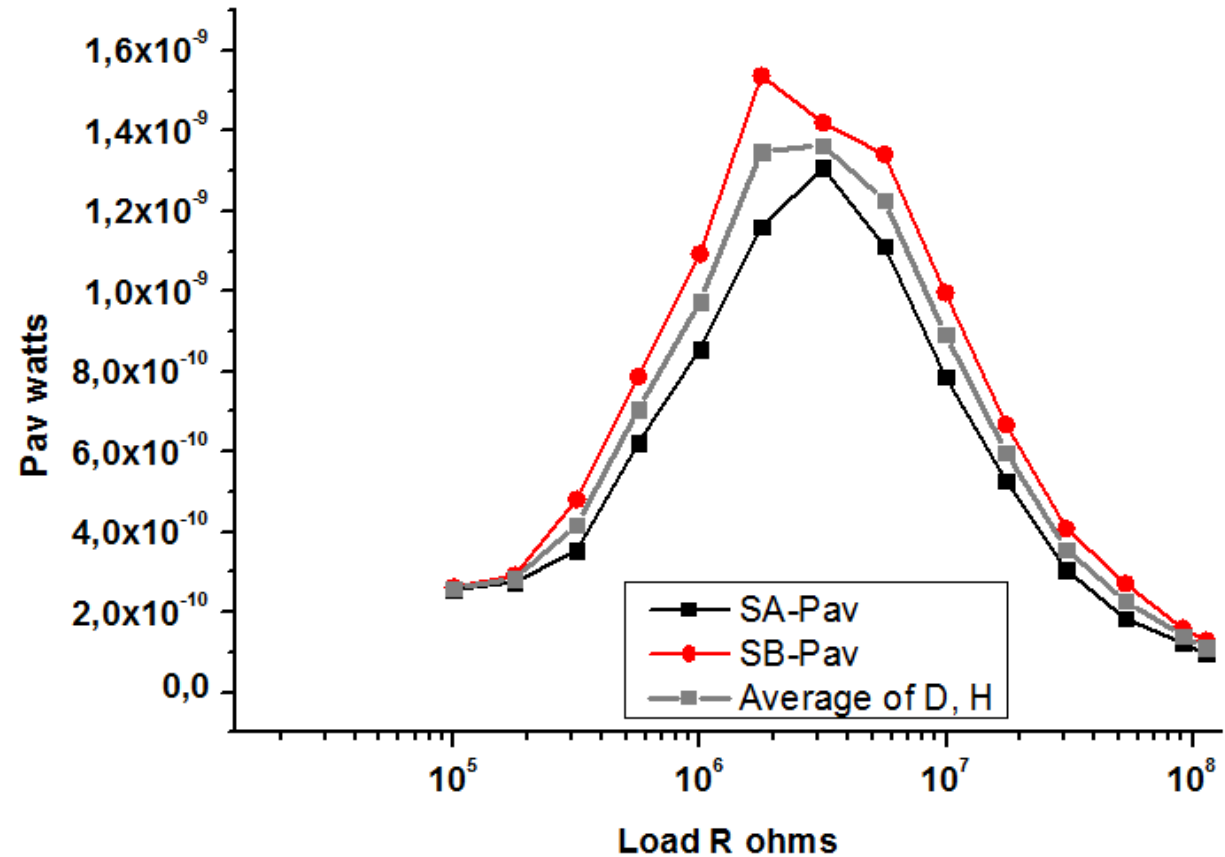
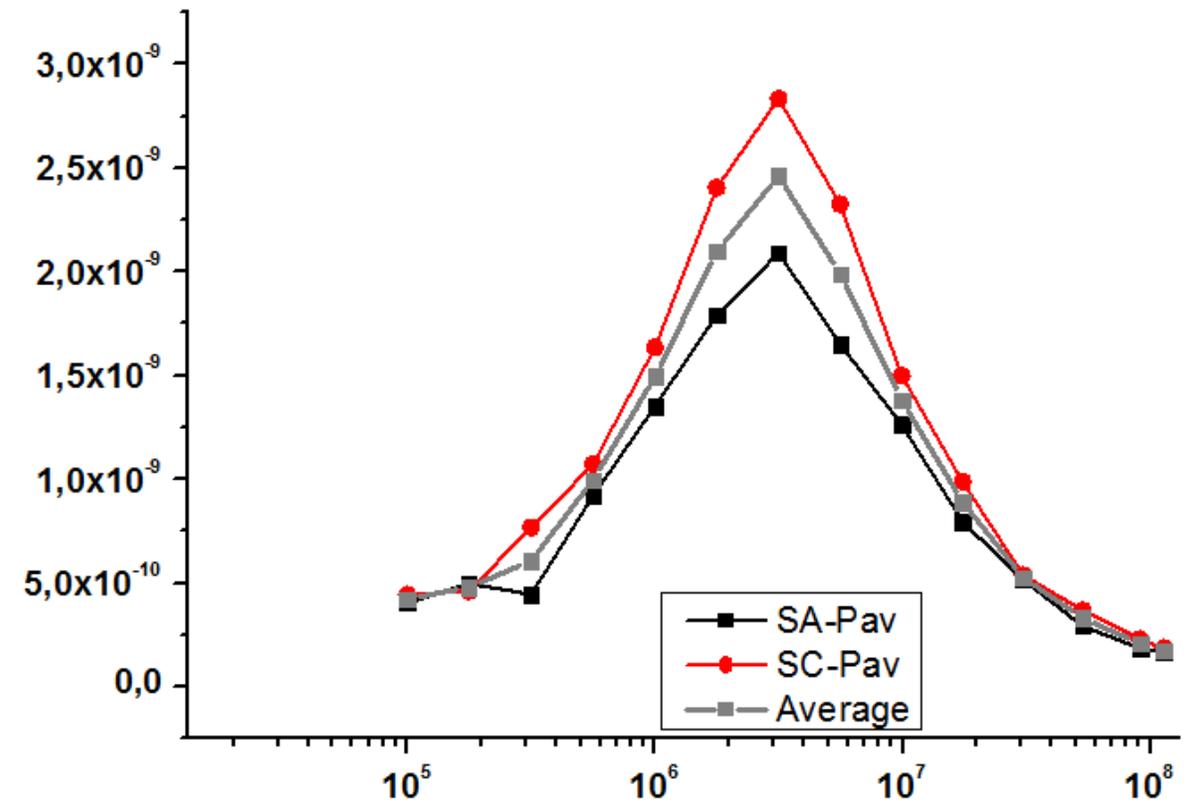

$P_{AV}$ delivered by the NGs subjected to a compression force at 5Hz: under 3N (left) and 6N (right)

# Nanowires on Ti/Au-Si : Growth using 30 mM Ammonia and Immersion in Liquid $N_2$ for 30 min

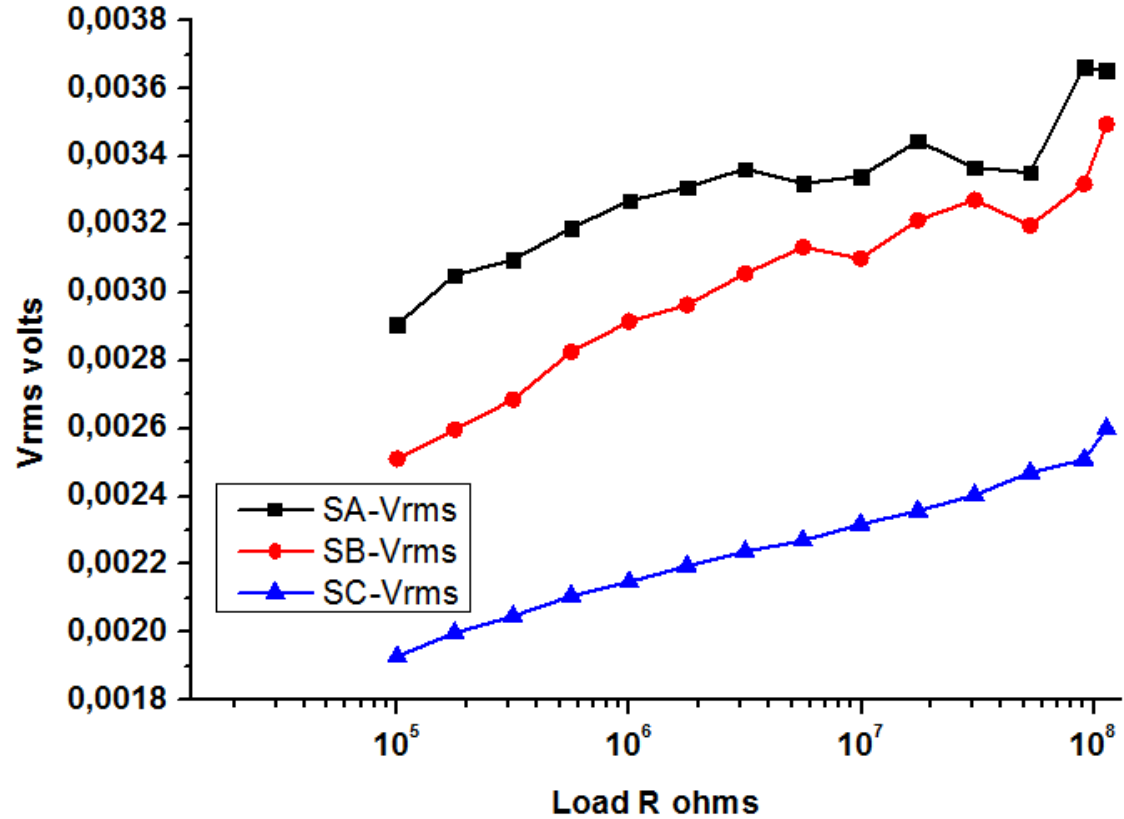

$V_{RMS}$ delivered by the NGs subjected to a 3N compression force at 5Hz

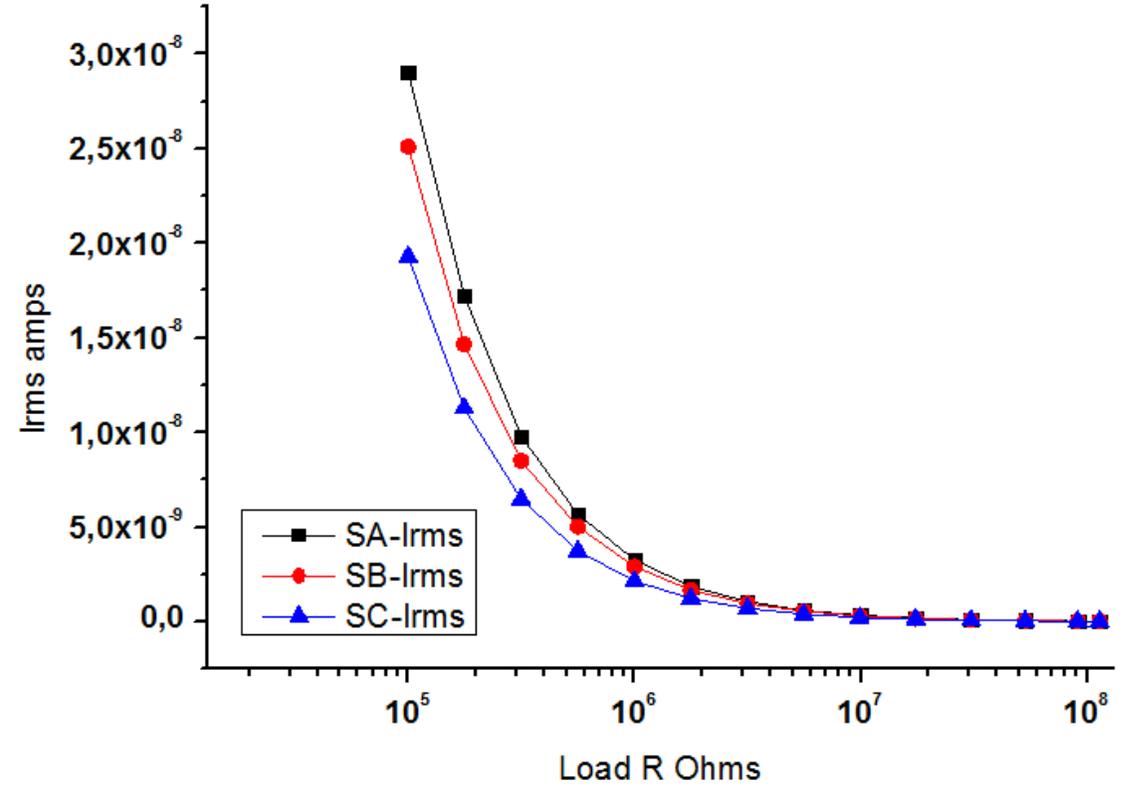

$I_{RMS}$ delivered by the NGs subjected to a 3N compression force at 5Hz

# Nanowires on Ti/Au-Si : Growth using 30 mM Ammonia and Immersion in Liquid N$_2$ for 30 min

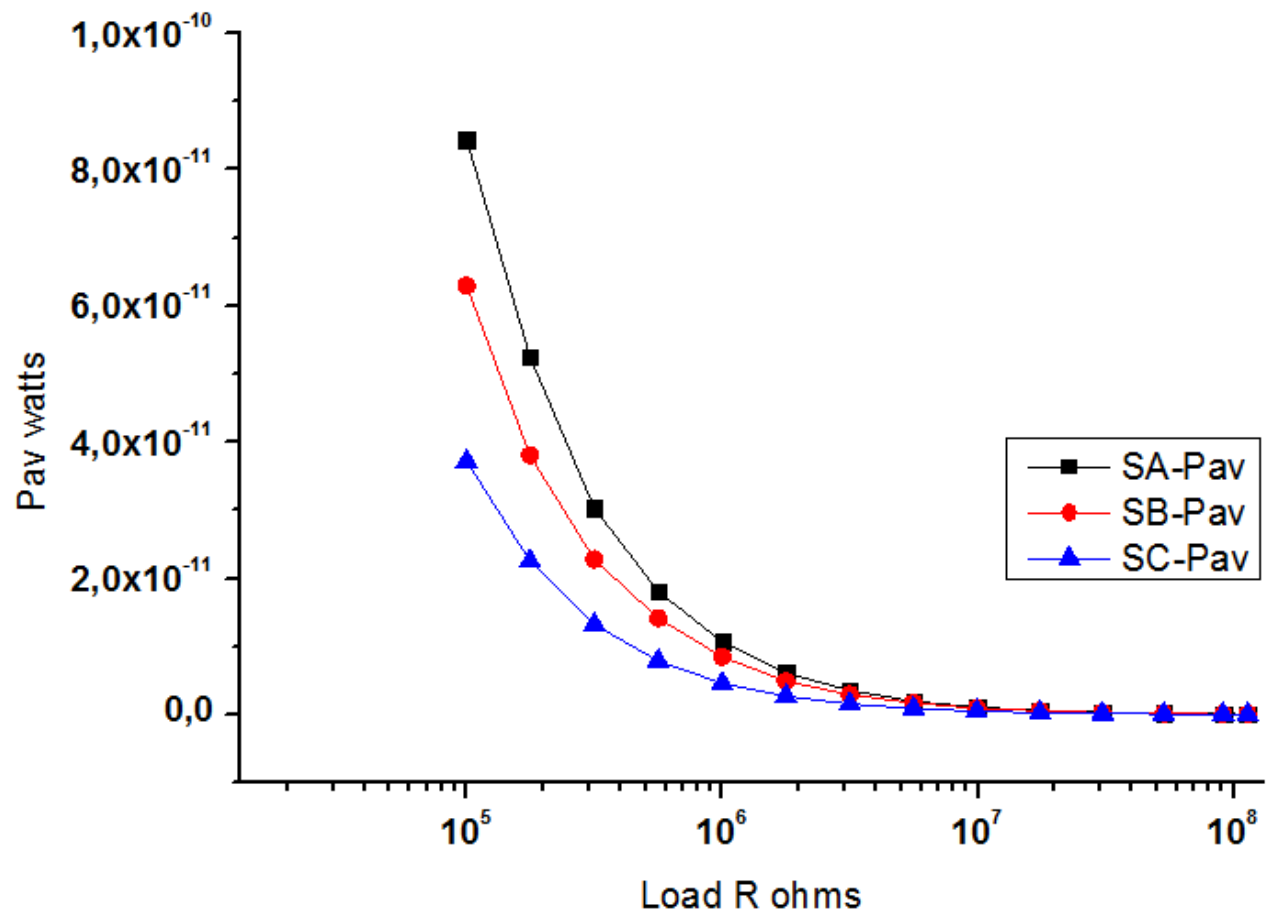

P$_{AV}$ delivered by the NGs subjected to a 3N compression force at 5Hz

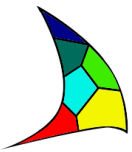
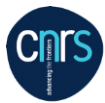

# 2. Effect of post-treatments:
# Thermal annealing and Immersion in Liquid Nitrogen

|  | at 3N Force | | | | at 6N Force | | |
| --- | --- | --- | --- | --- | --- | --- | --- |
| Sample Name | $V_{RMS}$ (V) | $I_{RMS}$ (A) | $P_{av}$ (W) | | $V_{RMS}$ (V) | $I_{RMS}$ (A) | $P_{av}$ (W) |
| Si_TiAu_30mM_growth_No annealing | 0.0038 | 2.36e-8 | 0.05e-9 | | 0.101 | 4.94e-8 | 1.19e-9 |
| Si_TiAu_30mM_growth_350°C **annealing** | 0.062 | 3.10e-8 | 0.54e-8 | | 0.141 | 5.87e-8 | 1.94e-9 |
| Si_TiAu_30mM_growth_450°C annealing | 0.086 | 3.80e-8 | 0.97e-9 | | 0.129 | 7.11e-8 | 2.35e-9 |
| | | | | | | | |
| Si_TiAu_30mM_growth_**LiquidN2_15 min** | 0.114 | 5.07e-8 | 1.38e-9 | | 0.141 | 6.39e-8 | 2.51e-9 |
| Si_TiAu_30mM_growth_LiquidN2_30 min | 0.0032 | 2.42e-8 | 0.05e-9 | | | | |

Maximum $V_{RMS}$, $I_{RMS}$ and $P_{AV}$ values of NGs prepared with 30mM ammonia concentration, and different post-treatments

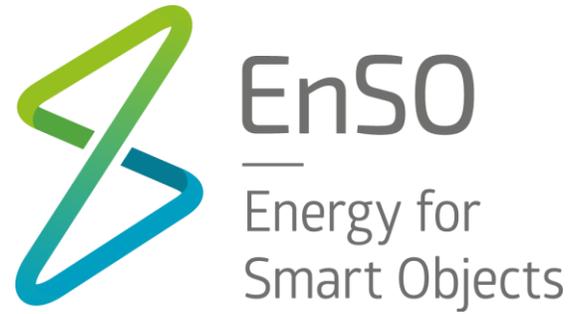 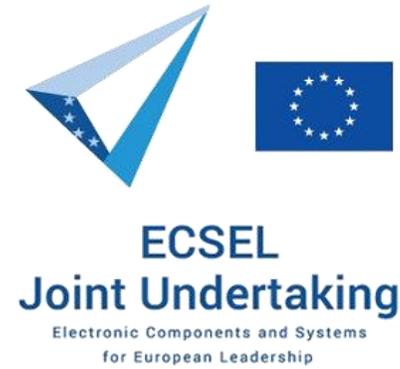

EnSO has been accepted for funding within the Electronic Components and Systems For European Leadership Joint Undertaking in collaboration with the European Union's H2020 Framework Programme (H2020/2014-2020) and National Authorities, under grant agreement n° 692482

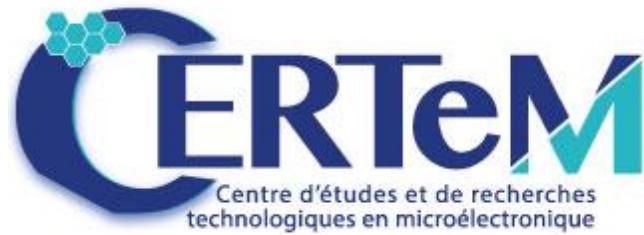

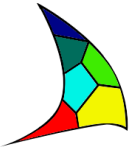 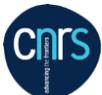 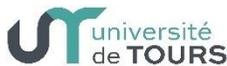 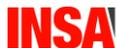